# The Fate of Primary Iron Sulfides in the CM1 Carbonaceous Chondrites: Effects of Advanced Aqueous Alteration on Primary Components

S. A. Singerling[1*], C. M. Corrigan[2], and A. J. Brearley[1]

[1] Department of Earth and Planetary Sciences, MSC-03 2040 1 University of New Mexico, Albuquerque, NM 87131, USA

[2] Department of Mineral Sciences, National Museum of Natural History, Smithsonian Institution, Washington, DC 20560, USA

*Corresponding author. Current institution: Department of Geosciences, Goethe University Frankfurt, Altenhoeferallee 1, 60438, Frankfurt am Main, Germany. E-mail: singerling@em.uni-frankfurt.de.

**Abstract**

We have carried out a SEM-EPMA-TEM study to determine the textures and compositions of relict primary iron sulfides and their alteration products in a suite of moderately to heavily-altered CM1 carbonaceous chondrites. We observed four textural groups of altered primary iron sulfides: 1) pentlandite+phyllosilicate (2P) grains, characterized by pentlandite with submicron lenses of phyllosilicates, 2) pyrrhotite+pentlandite+magnetite (PPM) grains, characterized by pyrrhotite-pentlandite exsolution textures with magnetite veining and secondary pentlandite, 3) pentlandite+serpentine (PS) grains, characterized by relict pentlandite exsolution, serpentine, and secondary pentlandite, and 4) pyrrhotite+pentlandite+magnetite+serpentine (PPMS) grains, characterized by features of both the PPM and PS grains.

We have determined that all four groups were initially primary iron sulfides, which formed from crystallization of immiscible sulfide melts within silicate chondrules in the solar nebula. The fact that such different alteration products could result from the same precursor sulfides within even the same meteorite sample further underscores the complexity of the aqueous alteration environment for the CM chondrites. The different alteration reactions for each textural group place constraints on the mechanisms and conditions of alteration with evidence for acidic environments, oxidizing environments, and changing fluid compositions (Ni-bearing and Si-Mg-bearing).

# 1. INTRODUCTION

Chondritic meteorites record a variety of primary and secondary processes and, as such, represent the best opportunity to study the earliest-formed materials in the solar nebula. This, in turn, provides insights into the evolution of the Solar System. The CM carbonaceous chondrites are of particular interest because their constituent primary components—chondrules, calcium-aluminium-rich inclusions (CAIs), and amoeboid olivine aggregates (AOAs) all embedded within matrix materials—have interacted with fluids (i.e., liquid and/or vapor water) to varying degrees. All of these components, especially the matrix, show evidence of aqueous alteration, indicated by the replacement of primary phases by a variety of secondary alteration products (i.e., phyllosilicates, tochilinite-cronstedtite intergrowths, iron oxides/hydroxides, and minor amounts of carbonates and sulfates (McSween 1987; Suttle et al. 2021)).

Aqueous alteration for CM chondrites took place from 10–245°C, based on oxygen isotopic studies of carbonates (e.g., Vacher et al. 2019a), and under neutral to alkaline conditions (e.g., DuFresne and Anders 1962; Brearley 2006a; Chizmadia and Brearley 2008; Vacher et al. 2019b). Studies of Mn-Cr ages and oxygen isotopes of carbonates show evidence for episodic alteration that lasted for ~10 Myr (De Leuw et al. 2009; Fujiya et al. 2012; Tyra et al. 2012). The CM chondrites vary in their extent of alteration by fluids from moderately (CM2 chondrites) to heavily altered (CM1 chondrites) (Browning et al. 1996; Zolensky et al. 1997; Rubin et al. 2007; Hewins et al. 2014; Howard et al. 2011; Howard et al. 2015; King et al. 2017; King et al. 2021). This variation in degree of alteration is significant, because it allows us to discern what changes the original primary components of the samples have undergone with different extents of alteration, and what they transform into as a result of interaction with fluids.

However, although the CM1 and CM2 chondrites likely stem from the same precursor material, the two may not necessarily be linked by progressive alteration (i.e., the CM1s are not more-altered products of the CM2s) as suggested by oxygen isotopic data and the presence or absence of certain phases. For example, dolomite occurs in the CM1s but rarely in the CM2s and tochilinite occurs in the CM2s but rarely in the CM1s (Clayton and Mayeda 1984; 1999; Howard et al. 2015; King et al. 2017). Instead, the CM1 chondrites may represent the products of aqueous alteration at higher temperatures (>100°C), for longer duration, or under higher water-rock ratios (Clayton and Mayeda 1984; McSween 1987; Clayton and Mayeda 1999; Tomeoka and Buseck 1985; Zolensky et al. 1989; 1997; Browning et al. 1996; Hanowski and Brearley 2001; Rubin et al. 2007)

The origin of iron sulfides in CM chondrites is controversial; both primary (solar nebula) and secondary (asteroidal parent body) origins have been proposed (e.g., Fuchs et al. 1973; Hanowski and Brearley 2001; Zolensky and Le 2003; Bullock et al. 2007; others below). Several studies have shown that coexisting pyrrhotite ($Fe_{1-x}S$) and pentlandite ($(Fe,Ni)_9S_8$) in CM2 chondrites could have formed by crystallization of monosulfide solid solution melts during the chondrule formation event(s) (Boctor et al. 2002; Brearley and Martinez 2010; Maldonado and Brearley 2011; Harries and Langenhorst 2013; Kimura et al. 2011; Hewins et al. 2014; Singerling and Brearley 2018). Other studies have argued that crystallization of sulfides during chondrule formation could describe textural and compositional features in CR, CV, and EH chondritic sulfides (e.g., Marrocchi and Libourel 2013; Schrader et al. 2015; Piani et al. 2016). Alternatively, formation of sulfides by sulfidization of Fe,Ni metal in the solar nebula has been proposed based on experimental works as well as observations in ordinary, CR, and CM chondrites (Zanda et al. 1995, Lauretta et al. 1996a; b; c; 1997; 1998; Schrader and Lauretta 2010;; Singerling and Brearley 2018). However, the primary sulfides undergo parent body alteration with increasing degrees of alteration of the bulk meteorite (Singerling and Brearley 2020).

We have previously described sulfides in CM2 chondrites in detail (Singerling and Brearley 2018). However, sulfides have only been described briefly in the highly-altered CM1 chondrites by Zolensky et al. (1997) where they documented the occurrence of two different kinds of sulfides: serpentine-sulfide aggregates and pyrrhotite-pentlandite-magnetite assemblages. They argued that these assemblages are secondary and formed on the parent body. However, the origins of and relationships between the sulfides in CM1 chondrites and the primary sulfides in CM2s has not yet been investigated. Based on the tendency for pyrrhotite to alter to magnetite or phyllosilicates in CM2 chondrites, one would expect that all primary pyrrhotite in CM1 chondrites would have been fully altered, implying that any pyrrhotite observed should be secondary in origin.

We have studied sulfides in CM1 chondrites to determine if they formed by primary or secondary processes based on a comparison with textures of primary sulfides as well as sulfides that have undergone partial alteration in CM2 chondrites. Given the evidence of alteration of primary sulfides in CM2 chondrites, our goal was to understand whether sulfides in CM1 chondrites are entirely the products of secondary alteration or whether remnants of primary sulfides are still present, even in these highly-altered meteorites. In addition, the assemblages have the potential to provide insights into the conditions of secondary alteration, and the mechanisms of alteration of primary sulfides.

## 2. METHODS

The textures and compositions of iron sulfide grains were studied in the following CM chondrites: ALH 83100, ALH 84029, ALH 84034, ALH 84049, LAP 031166, and MET 01073. Data were obtained from the following polished thin sections (PTSs): ALH 83100,12; ALH 84029,40; ALH 84034,9; ALH 84049,10; LAP 031166,13; and MET 01073,9. The Allan Hills (ALH/A) samples listed are all one pairing group (MacPherson 1985a; MacPherson 1985b; Mason 1986). All PTSs are part of the U.S. Antarctic Meteorite Collection and were obtained from NASA's Astromaterials Acquisition and Curation Office.

All samples except MET 01073 are classified as CM1/2 chondrites according to the Meteoritical Bulletin; however, we argue that they are more appropriately considered as CM1 chondrites. The near complete hydration of ALH 84034 and ALH 84049 (Llorca and Brearley 1992; Tyra 2013) implies that these meteorites, and, by extension members of their pairing group, (i.e., ALH 84029), are CM1 chondrites. Additionally, studies of the modal abundances, specifically the proportions of anhydrous silicates to total phyllosilicates, in ALH 83100 and LAP 031166 demonstrate that these meteorites have experienced similar degrees of aqueous alteration (Howard et al. 2011; King et al. 2017) again arguing for a CM1 classification for all the samples.

For back-scattered electron (BSE) imaging of the textures and energy dispersive X-ray spectrometry (EDS) analyses, we used an FEI Nova NanoSEM 600 at the National Museum of Natural History in the Mineral Sciences Department using the following operating conditions: 6 mm working distance, 15 kV accelerating voltage, and 1.4 nA beam current. Preparation of FIB sections for use on the Transmission Electron Microscope (TEM) was performed on a FEI Quanta 3D DualBeam® Field Emission Gun Scanning Electron Microscope/Focused Ion Beam (FEGSEM/FIB) at the University of New Mexico in the Department of Earth and Planetary

Sciences. The focused ion beam (FIB) sections, protected from beam damage using an ion-beam deposited platinum strip, were removed from the thin section using the *in situ* lift out technique using an Omniprobe 200 micromanipulator. After extraction from the thin section, the FIB samples were mounted onto Cu TEM half grids. Samples were then milled to electron transparency. Extraction of the FIB samples was performed at an ion beam accelerating voltage of 30 kV with a beam current ranging between 1–5 nA. Milling to electron transparency was also carried out at an ion beam accelerating voltage of 30 kV, with beam currents decreasing from 0.5 nA–50 pA at the final stage.

The major and minor element compositions of the sulfides were obtained using wavelength dispersive spectrometry (WDS) on a JEOL 8200 Electron Probe Microanalyzer (EPMA) in the Institute of Meteoritics, University of New Mexico. Operating conditions were 15 kV accelerating voltage, 20 nA beam current, and a beam size of <1–5 μm depending on the specific phase being analyzed and the overall size of the grain. Elements analyzed, the crystals they were measured on, count times, detections limits, and standards used are summarized in the appendix (Table A1). Appropriate corrections were made for elements whose peaks interfere with one other (i.e., Fe and Co) using the Probe for EPMA (PFE) software (Donovan et al. 1993). Standard ZAF corrections were applied to the data within the Probe for EPMA software. Note that the size of sulfide grains featured in this study (i.e., >10 μm) was largely chosen based on the spot size of EPMA analyses.

The JEOL 2010F Scanning/Transmission Electron Microscope (S/TEM) in the Department of Earth and Planetary Sciences at University of New Mexico was operated at 200 kV to obtain high-angle annular dark-field (HAADF) STEM images, selected area electron diffraction patterns, and EDS X-ray spot analyses and maps. EDS data were obtained using an Oxford Instruments AZtecEnergy EDS system coupled to an Oxford X-Max 80 mm$^2$ silicon drift detector. The JEOL NEOARM 200CF aberration corrected S/TEM in the Nanomaterials Characterization Facility at the University of New Mexico was operated at 200 kV to obtain high resolution (HR) TEM bright field images and EDS X-ray maps. EDS data were obtained using an Oxford Instruments AZtecEnergy EDS system coupled to two JEOL 100 mm$^2$ silicon drift detectors.

## 3. RESULTS

All of the CM1 chondrites studied are completely hydrated. The samples are dominated by serpentine with lesser amounts of sulfides and carbonates. We searched the surface area of each thin section including both matrix and pseudomorphed chondrules. We define a pseudomorphed chondrule as a roughly circular feature at least 100 μm in diameter consisting of serpentine that is texturally and compositionally distinct (i.e., lower Z contrast) from the surrounding matrix and containing opaques on the chondrule rims or within olivine pseudomorphs in the chondrule interiors (Fig. A1b). We limited our study of altered primary iron sulfides to coarse-grained phases (i.e., >10 μm in size).

Our observations yielded several common textural groups, which have characteristics that suggest that they may represent primary sulfides which have undergone aqueous alteration. These include the following:
1) Pentlandite+phyllosilicate (2P) grains
2) Pyrrhotite+pentlandite+magnetite (PPM) grains
3) Pentlandite+serpentine (PS) grains
4) Pyrrhotite+pentlandite+magnetite+serpentine (PPMS) grains

These grains were considered altered based on textural evidence (e.g., presence of pores or reaction fronts, pseudomorphic replacement, etc.). Table 1 summarizes the abundance, range in size, spatial occurrence, and exsolution textures (i.e., patches, blades, lamellae, or rods) of each textural type by sample. Figure 1 shows examples of these textural types, whereas Figure A2 shows examples of the pentlandite exsolution textures.

**Table 1.** Textural groups, number of grains identified, sizes, spatial occurrence, and exsolution textures of altered primary sulfide grains in the CM1 chondrites studied.

| Sample | Textural group | Number of grains | Size range (μm) | Spatial occurrence | Exsolution textures |
|---|---|---|---|---|---|
| ALH 84029 | 2P | 5 | 18−85 | Matrix | N/A |
| | PPM | 3 | 30−110 | Matrix | p,b,r |
| | PS | 2 | 30−40 | Chondrule, matrix | p,b |
| | PPMS | 2 | 60−125 | Mx | p,l |
| ALH 84034 | 2P | 4 | >10−25 | Matrix | N/A |
| | PPM | 2 | 25−35 | Matrix | p |
| | PS | 4 | 15−60 | Chondrule, matrix | p,b,l |
| | PPMS | 1 | 40 | Matrix | p |
| ALH 84049 | 2P | 3 | 55−75 | Matrix | N/A |
| | PPM | 4 | 30−55 | Chondrule, matrix | p |
| | PS | 3 | 35−40 | Matrix | p,b |
| | PPMS | 3 | 35−50 | Matrix | p |
| ALH 83100 | PPM | 6 | 15−90 | Chondrule, matrix | p,b,r |
| LAP 031166 | 2P | 1 | 35 | Matrix | N/A |
| | PS | 1 | 25 | Matrix | p,r |
| MET 01073 | 2P | 1 | 35 | Matrix | N/A |
| | PS | 2 | 30−75 | Chondrule, matrix | p,r |

p = patches, b = blades, l = lamellae, r = rods, N/A = no textures present
Note that >10 μm is listed in the size range column owing to the fact that 10 μm was the minimum grain size studied in this work.

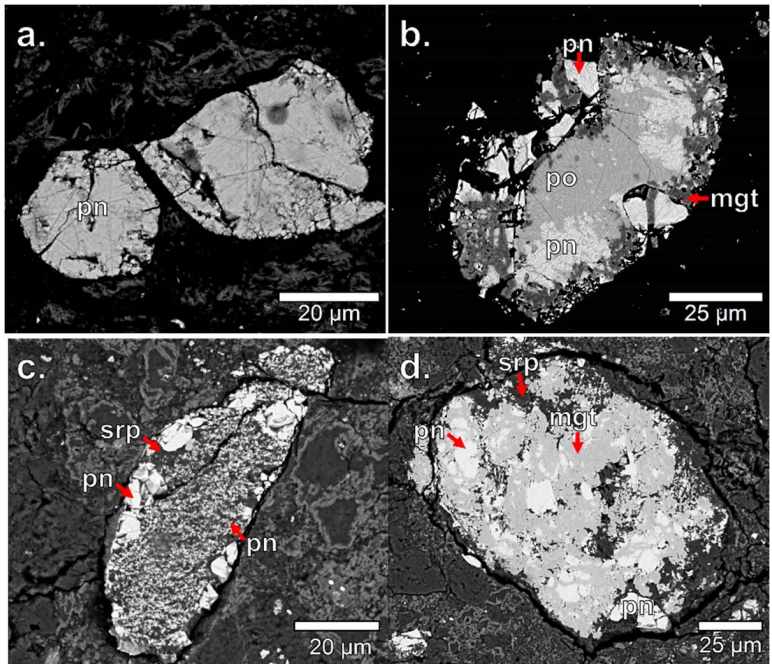

**Figure 1.** BSE images showing examples of the textural groups of altered primary sulfides in CM1 chondrites—(a) pentlandite+phyllosilicate (2P) grain, (b) pyrrhotite+pentlandite+magnetite (PPM) grain, (c) pentlandite+serpentine (PS) grain, and (d) pyrrhotite+pentlandite+magnetite+serpentine (PPMS) grain.

### 3.1 Pentlandite+phyllosilicate (2P) grains

The porous pentlandite (2P) grains (Fig. 2) are found in all samples studied except ALH 83100 and only occur as isolated grains in the matrix. The grains range in size from >10–85 µm and are anhedral to euhedral, with the latter having hexagonal forms. The 2P grains consist of an Fe-Ni sulfide with a lamellar texture that is characterized by variations in Z contrast in HAADF STEM images. Based on SAED patterns (Fig. 2d), as well as EDS spot analyses and X-ray maps (Fig. 2f), this phase is pentlandite, and the differing Z contrast between lamellae are best explained as twinning. In addition, the pentlandite contains abundant crystallographically-oriented features with lower Z contrast than the host pentlandite.
HRTEM imaging (Fig. 2c) and EDS mapping (Fig. 2f) show that the pore-like features are filled with Mg-rich phyllosilicates. The phyllosilicates have a heterogeneous distribution within a given grain and between grains. These phyllosilicate lenses range in size from <1 µm to a few microns; the vast majority are submicron and range in shape from round to ellipsoidal to linear. The phyllosilicate lenses have lengths varying from <5 to 625 nm in the direction parallel to the pentlandite lamellae and widths varying from <5 to 150 nm in the direction perpendicular to the lamellae. The HAADF STEM images (Fig. 2d−e) clearly show the crystallographic orientation of the lenses is most often parallel to the twinning interface. However, less well-developed lenses of phyllosilicates oriented at 60-120° to the twinning (see yellow arrows in Fig. 2e) also occur.

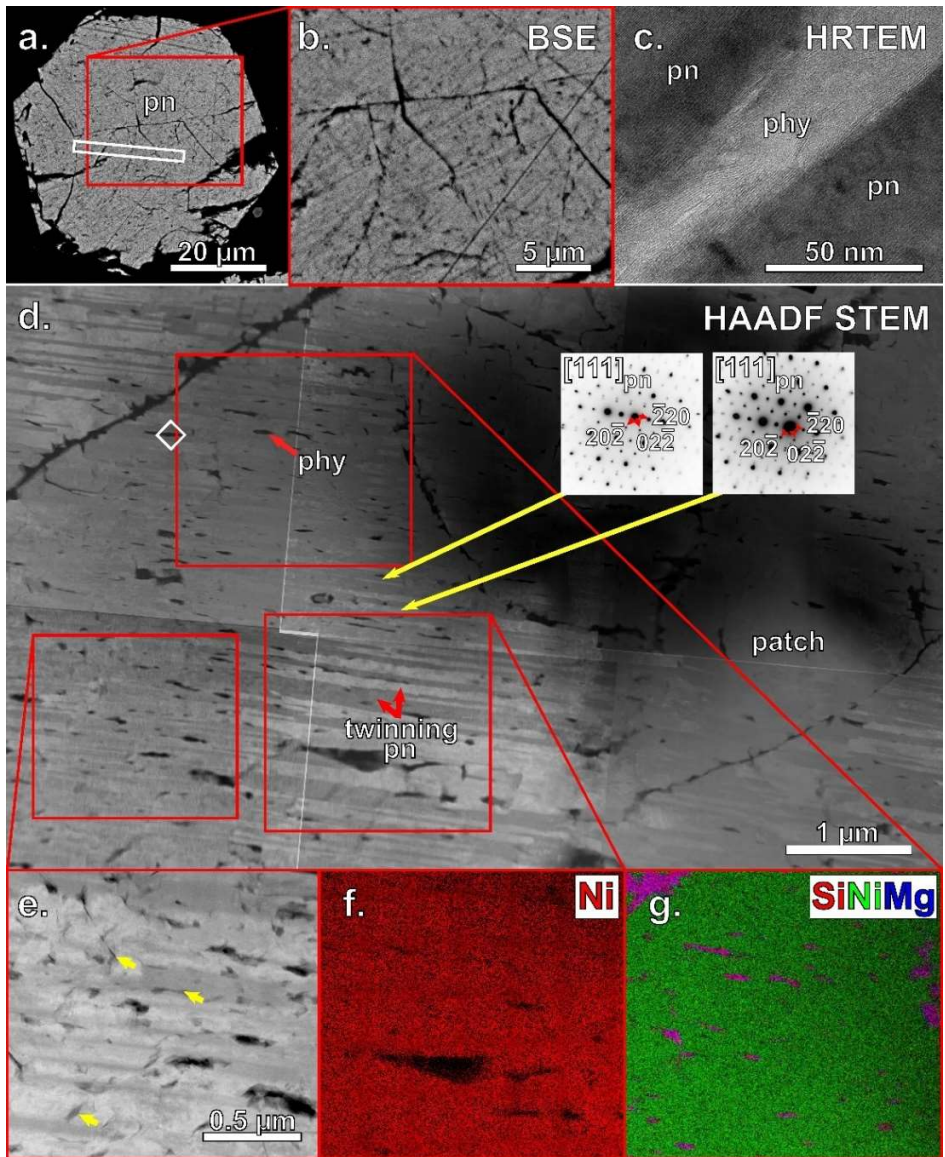

**Figure 2.** SEM, TEM, and STEM images of a 2P grain from CM1 ALH 84049. BSE images show (a) the euhedral pentlandite grain with the location of the FIB section extraction and (b) a high magnification image of the grain. A HRTEM bright-field image (c) of one of the pore-like features illustrates that it is actually filled with phyllosilicates. (d) includes a HAADF STEM mosaic of the FIB section as well as SAED patterns of pentlandite that shows lamellar variations in Z contrast that are due to twinning. The small, white rectangle indicates the location of (c). A high magnification HAADF STEM image (e) shows phyllosilicate-filled features with three different orientations highlighted by the yellow arrows. The fibrous texture is consistent with phyllosilicates. The EDS X-ray map for Ni (f) shows the lack of compositional variations in Ni between lamellae while low and high Z contrast can be seen in HAADF images, which is consistent with EDS spot analyses and SAED patterns showing both have a pentlandite composition and structure. The EDS X-ray map for Si, Ni, and Mg (RGB) (g) shows the presence of Si,Mg-bearing phases in the low-Z contrast features (g), which are consistent with observations from HRTEM imaging (c) showing the presence of phyllosilicates. Pn = pentlandite, phy = phyllosilicates.

These are parallel to $\{112\}_{pn}$, which are planes of high packing density with (Fe+Ni):S of 9.5:7 for one unit cell. The lenses in the patches, on the other hand, are oriented parallel to $\{220\}_{pn}$, which are also planes of high density packing of atoms but with (Fe+Ni):S of 7.5:4 for one unit cell. In both the pentlandite lamellae and the pentlandite patch, the phyllosilicates occur along planes with greater Fe and Ni than S atoms.

Table 2 presents individual representative EPMA analyses of pyrrhotite and pentlandite in the 2P, PPM, and PS grains, and are discussed in detail below. See the appendix (Tables A2 and A3) for the complete data. Compositional data for the 2P grains are presented graphically in Figure 3a. Analyses from the ALH pairing group (ALH 84029, 84034, and 84049) all largely overlap, whereas analyses from LAP 031166 are notably lower in their Co contents. The 2P grains range in composition from 30.3–32.6 wt. % Ni and 1.6–3.1 wt. % Co. While the Ni contents are typical of CM2 chondrite pentlandite, the Co contents are higher; in Singerling and Brearley (2018), we observed a maximum Co content of 1.3 wt. % for primary pentlandite in CM2 chondrites.

### 3.2 Pyrrhotite+pentlandite+magnetite (PPM) grains

The pyrrhotite+pentlandite+ magnetite (PPM) grains (Fig. 4) are found in all samples studied, except LAP 031166 and MET 01073, and occur within pseudomorphed chondrules as well as in the matrix. The grains range in size from 30–110 µm and are anhedral. These grains have textures consisting of pyrrhotite with patches, blades, and rods of pentlandite, resembling those in primary pyrrhotite-pentlandite grains in CM2 chondrites, as well as magnetite veining. Two dominant textures of pentlandite are observed: patch-textured and blade-textured. The former occurs along the periphery of the grains and is euhedral to subhedral (Fig. 4a). The latter can occur throughout the grain, both near the periphery and in the interior, and is often composed of subparallel blades (Fig. 4b).

The proportions of pyrrhotite, pentlandite, and magnetite vary from grain to grain both within the same sample and between the different meteorites. In the least-altered grains, magnetite veining and pentlandite are limited to the outer portions of the grain, and pyrrhotite is still present in major amounts. In more-altered grains, magnetite veining crosscuts the grain entirely, pentlandite is present in greater proportions, and pyrrhotite is only present in small amounts.

Magnetite is nearly always in direct contact with blade-textured pentlandite, although there are cases where it is in contact with pyrrhotite (Fig. 4b). Evidence of a crystallographic orientation relationship between magnetite and either pyrrhotite or pentlandite is not apparent from SEM images. Instead, the magnetite veins often follow the grain boundaries in the least-altered examples. The blade-textured pentlandite in contact with pyrrhotite and magnetite, however, does appear to be crystallographically oriented relative to the pyrrhotite based on SEM observations (Fig. 4b).

TEM analyses were performed on a FIB section extracted from a representative less-altered PPM grain in ALH 83100 that contains pyrrhotite, pentlandite, and magnetite. As also seen in SEM images, a crystallographic orientation of the pentlandite is apparent with respect to the pyrrhotite. In some areas of the grain, magnetite is in direct contact with pyrrhotite. The TEM studies show that the magnetite veining consists of a polycrystalline, monomineralic aggregate with randomly-oriented, blocky grains a few hundred nanometers in size (Fig. 4c, 4e). Pentlandite occurs as elongate grains parallel to one another (Fig. 4c, 4e). They range in width from 20–600 nm and in length from 270 nm to a few microns. Where magnetite is in contact

**Table 2.** Representative EPMA spot analyses (in wt. %) of individual phases in 2P, PPM, and PS grains in CM1 chondrites.

| Textural group | Meteorite | Grain | Phase | P | S | Cr | Fe | Co | Ni | Total |
|---|---|---|---|---|---|---|---|---|---|---|
| 2P | ALH 84029 | S13 | Pn | bdl | 32.62 | bdl | 30.55 | 2.86 | 31.21 | 97.2 |
| | | | | bdl | 32.61 | bdl | 29.77 | 2.80 | 31.12 | 96.3 |
| | | | | bdl | 32.68 | bdl | 30.10 | 2.80 | 31.35 | 97.0 |
| | | | | bdl | 32.75 | bdl | 29.76 | 2.61 | 31.09 | 96.2 |
| | | | | bdl | 31.48 | bdl | 29.45 | 1.56 | 32.57 | 95.1 |
| | | | | bdl | 32.75 | bdl | 30.09 | 2.85 | 31.44 | 97.2 |
| | | | | bdl | 31.80 | bdl | 30.03 | 2.65 | 30.70 | 95.2 |
| | | | | bdl | 32.71 | bdl | 30.10 | 2.71 | 31.15 | 96.7 |
| | LAP 031166 | S9 | Pn | bdl | 32.62 | 0.04 | 30.76 | 2.11 | 32.04 | 97.6 |
| | | | | bdl | 32.33 | bdl | 30.45 | 2.07 | 31.35 | 96.2 |
| | | | | bdl | 32.21 | 0.05 | 30.24 | 2.07 | 31.85 | 96.4 |
| | | | | bdl | 32.47 | bdl | 30.45 | 2.21 | 31.98 | 97.1 |
| | | | | bdl | 32.65 | bdl | 30.66 | 2.25 | 32.32 | 97.9 |
| | | | | bdl | 32.88 | 0.05 | 30.82 | 2.21 | 32.23 | 98.2 |
| | | | | bdl | 32.28 | bdl | 30.99 | 2.07 | 31.70 | 97.1 |
| | MET 01073 | S5 | Pn | bdl | 32.40 | bdl | 33.20 | 1.06 | 31.78 | 98.5 |
| | | | | bdl | 32.62 | bdl | 32.59 | 0.92 | 32.34 | 98.5 |
| | | | | bdl | 31.14 | 0.05 | 34.33 | 1.15 | 29.87 | 96.5 |
| PPM | ALH 84049 | S6 | Pn p | bdl | 31.52 | bdl | 30.45 | 1.67 | 33.16 | 96.8 |
| | | | Pn b | bdl | 32.36 | bdl | 32.97 | 0.53 | 31.11 | 97.0 |
| | | | Po | bdl | 37.46 | bdl | 56.91 | bdl | 2.24 | 96.6 |
| | LAP 03116 | S5 | Pn p | bdl | 32.72 | 0.04 | 35.90 | 1.01 | 26.03 | 95.7 |
| | | | Po | 0.02 | 37.97 | 0.05 | 58.69 | 0.08 | 1.71 | 98.5 |
| | | | | bdl | 37.88 | 0.04 | 58.56 | 0.07 | 1.68 | 98.2 |
| | | | | bdl | 37.97 | 0.06 | 56.92 | 0.12 | 3.05 | 98.1 |
| | | | | bdl | 37.87 | 0.08 | 57.52 | 0.15 | 2.86 | 98.5 |
| | | | | bdl | 37.78 | 0.07 | 58.57 | 0.10 | 1.93 | 98.4 |
| | | | | bdl | 38.00 | 0.07 | 58.67 | 0.07 | 1.73 | 98.5 |
| PS | ALH 84049 | S3 | Pn c | bdl | 32.31 | bdl | 30.99 | 0.84 | 32.50 | 96.7 |
| | | | | bdl | 32.28 | bdl | 31.48 | 0.87 | 32.47 | 97.1 |

pn = pentlandite, p = patch-textured, b = blade-textured, c = coarse-grained, po = pyrrhotite, bdl = below detection limit with elongate pentlandite, the phases are oriented with $(220)_{mgt}//(511)_{pn}$, with the pentlandite elongation parallel to $\{111\}_{pn}$ (Fig. 4c). A minor amount of Fe,Mg serpentine, identified using EDS spot analyses, is present as small grains (~70 nm in size) that occur interstitially between pentlandite and magnetite (Fig. 4d). Additionally, rare grains of an Fe,S,O-bearing phase (possibly tochilinite) are present between the pentlandite and magnetite (Fig. 4d).

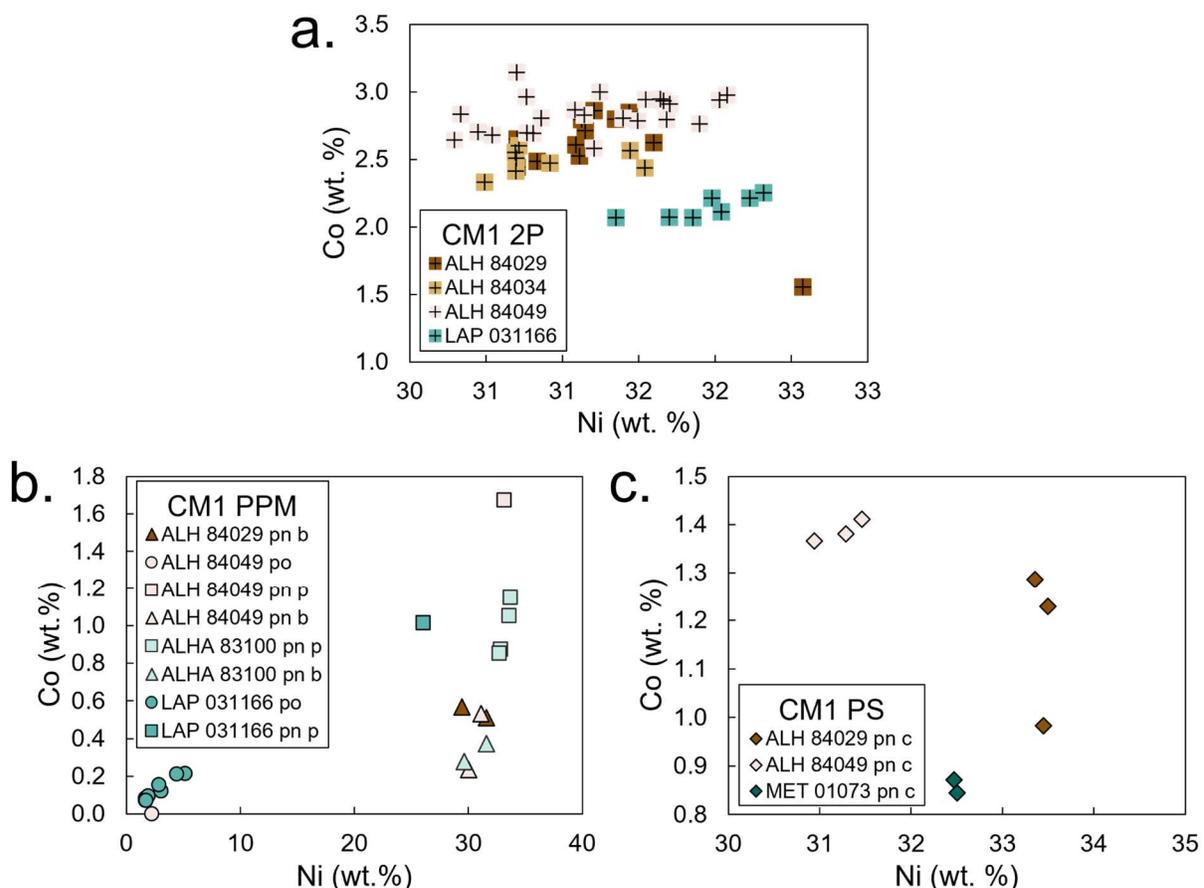

**Figure 3.** Co versus Ni plots (in wt. %) for individual spot analyses of (a) pentlandite in CM1 2P grains, (b) pyrrhotite and pentlandite in CM1 PPM grains, and (c) euhedral to subhedral pentlandite along grain boundaries in CM1 PS grains. Note the differences in scale for both axes between all panels. The points are color-coded according to the sample they were observed in. The shape of the symbol corresponds to the phase as defined in the legends. Error bars are smaller than the size of the symbols. Po = pyrrhotite, pn p = patch-textured pentlandite, pn b = blade-textured pentlandite, pn c = coarse-grained pentlandite.

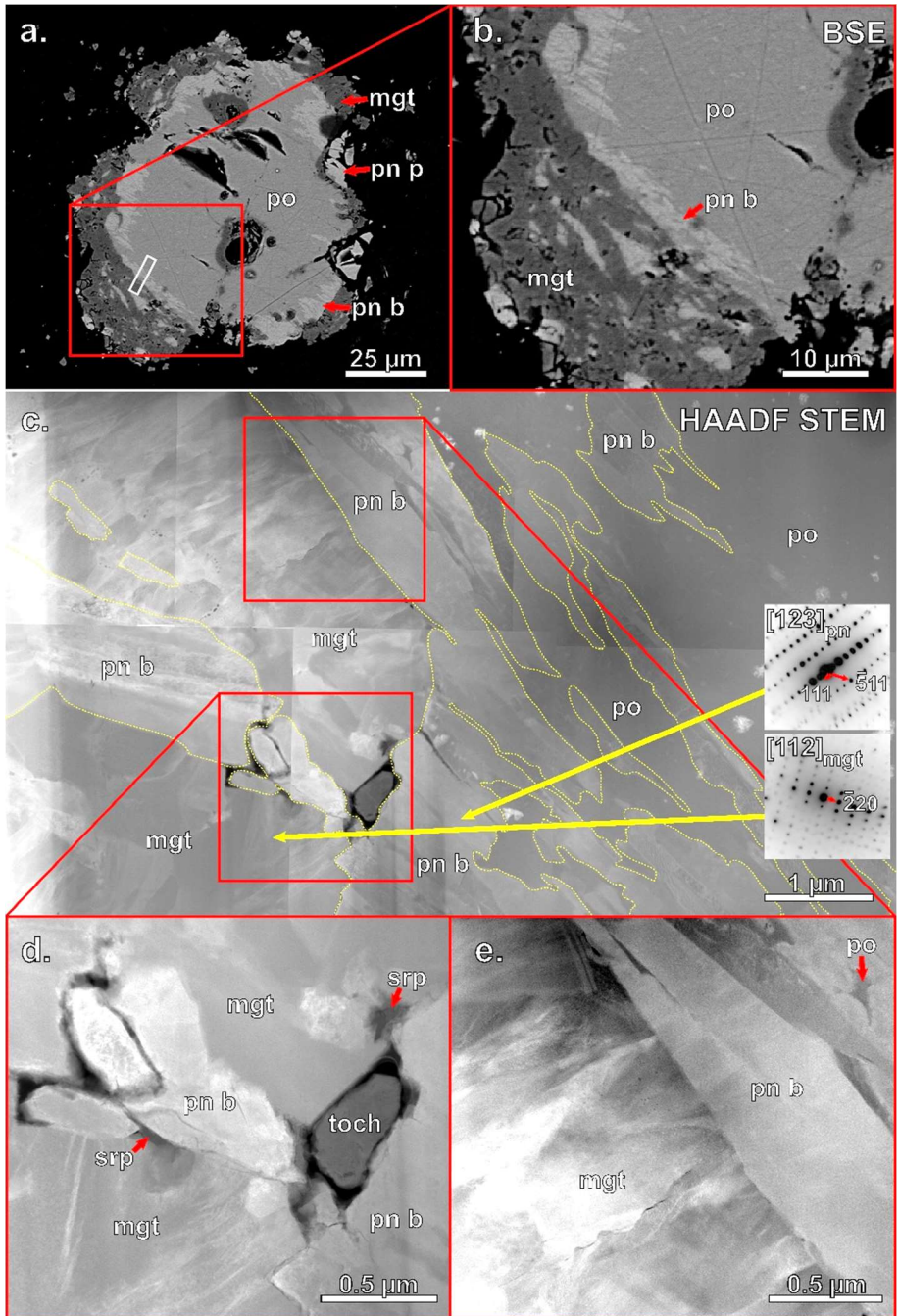

**Figure 4.** SEM and STEM images of a PPM grain located in CM1 ALH 83100. BSE images show (a) the overall grain, containing large proportions of pyrrhotite with magnetite veining largely limited to the grain boundaries, with the location of the FIB section extraction and (b) a high magnification image of the grain illustrating the crystallographic orientation of the blade-textured pentlandite with respect to the pyrrhotite. (c) includes a HAADF STEM mosaic of the FIB section as well as SAED patterns of crystallographically-oriented grains of magnetite and blade-textured pentlandite (outlined in yellow). High magnification HAADF STEM images show (d) a complex phase assemblage and (e) the contact for the magnetite and blade-textured pentlandite boundary. Po= pyrrhotite, pn = pentlandite, p = patch-textured, b = blade-textured, mgt = magnetite, srp = serpentine, toch = tochilinite-like.

The PPM grain composition data are summarized in Table 2 and are presented graphically in Figure 3b. We obtained analyses for pyrrhotite and two textural forms of pentlandite, patch and blade. Among all meteorites, the PPM grain pyrrhotite contains 1.7–5.2 wt. % Ni and has Co contents from below detection limit (<0.03) to 0.2 wt. %; the PPM grain patch-textured pentlandite ranges in Ni content from 26.0–33.7 wt. % and in Co content from 0.6–1.7 wt. %; and the PPM grain blade-textured pentlandite ranges in Ni content from 29.6–31.6 wt. % and in Co content from 0.2–0.5 wt. %.

Comparing the compositions by meteorite, we find that the ALH pairing group members (ALH/A 83100, 84029, and 84049) have similar Ni and Co compositions for pentlandite; the Co contents vary within a given meteorite. For example, among the samples that have both textural forms of pentlandite, ALH 83100 shows the smallest range in Co contents (0.3–1.2 wt. %), whereas ALH 84049 shows the greatest range (0.2–1.7 wt. %). Pentlandite from LAP 031166 has lower Ni contents compared to the ALH samples though its Co contents are similar. Pyrrhotite from LAP 031166 has similar Ni contents as pyrrhotite from ALH 84049 but has higher Co contents.

Differences between meteorite samples, however, are less significant than differences between textural types. While the Ni contents overlap between the two textural forms of pentlandite, the Co contents form two distinct groups: a low Co (<0.6 wt. %) and a moderate Co group (>0.8 wt. %). The low and moderate Co groups are composed of the blade-textured and patch-textured pentlandite, respectively.

### 3.3 Pentlandite+serpentine (PS) grains

The third group of altered primary sulfide grains are the pentlandite+serpentine (PS) grains (Fig. 5). These grains are similar to the pyrrhotite-pentlandite intergrowth (PPI) grains in Singerling and Brearley (2018). The PPI grains contain pyrrhotite with smaller amounts of pentlandite exsolution, which occurs as patches along the periphery of the grains and/or rods and blades in the interior of the grain. The PS grains also contain pentlandite around the periphery of the grains. Unlike the PPI grains, however, the PS grains do not contain any pyrrhotite but, instead, contain serpentine and pentlandite (both coarse-grained along grain boundaries and fine-grained in grain interiors). The PS grains are found in all samples studied except ALH 83100, occur in relict chondrules as well as the matrix, range in size from 15 μm to 130 μm, and are anhedral.

The PS grains have variable amounts of serpentine and pentlandite from grain to grain both within the same sample and between the different meteorites; however, the PS grains in MET 01073 have notably higher proportions of serpentine over pentlandite in the two PS grains observed. Within a single grain, the distribution of serpentine and fine-grained pentlandite is heterogeneous; some parts of the grain contain regions of serpentine without any fine-grained pentlandite, while other areas have more equal proportions of the two phases (Fig. 5a).

TEM analyses were made on a FIB section extracted from a representative PS grain in ALH 84049. The FIB section includes all phases identified by SEM: serpentine and pentlandite.

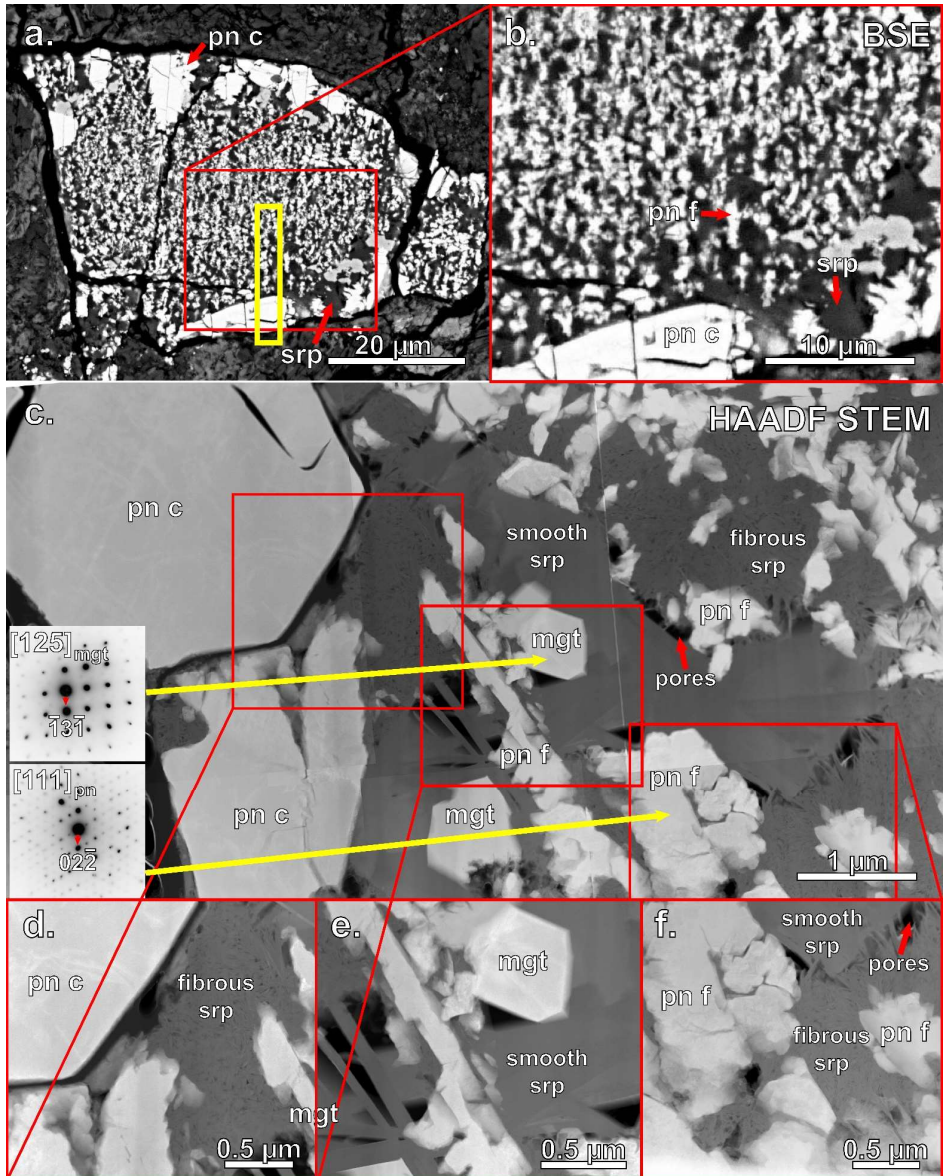

**Figure 5.** SEM and STEM images of a PS grain located in CM1 ALH 84049. BSE images show (a) the overall grain along with the location of the FIB section extraction and (b) a high magnification image of the grain illustrating the somewhat heterogeneous distribution of serpentine and fine-grained pentlandite. (c) includes a HAADF STEM mosaic of the FIB section as well as SAED patterns of crystallographically-oriented grains of magnetite and fine-grained pentlandite. High magnification HAADF STEM images show phase boundaries between (d) pentlandite and serpentine, (e) magnetite, pentlandite, and smooth serpentine, and (f) pentlandite and smooth and fibrous serpentines. Pn = pentlandite, c = coarse grained, f = fine grained, mgt = magnetite, srp = serpentine.

TEM analyses reveal the presence of coarse- and fine-grained pentlandite, two textural forms of serpentine, and euhedral magnetite grains. Additionally, porosity is observed predominantly in association with contacts between the two textural forms of serpentine.

The coarse-grained pentlandite (5–15 µm) consists of well-facetted, euhedral single crystals, which show limited evidence of secondary alteration or subhedral grains. Some coarse-grained pentlandite shows evidence of secondary replacement, indicated locally by jagged, irregular edges (Fig. 5a, 5c). The fine-grained pentlandite (30 nm–1.25 µm) crystals, embedded within serpentine, vary in morphology from rod-shaped to flower-like clusters (Fig. 5e–f), as well as highly irregularly-shaped grains. EDS analyses of the two textural types show that the compositions of the two agree within error.

Serpentine, identified by electron diffraction and EDS analyses, displays two textural forms: smooth (Fig. 5e) and fibrous (Fig. 5d). Based on electron diffraction analyses, the smooth serpentine consists of single crystals, whereas the fibrous serpentine consists of numerous submicron-sized (less than 20 nm in width and 300 nm in length) fibers in random orientations. The smooth serpentine is associated with magnetite and both fine- and coarse-grained pentlandite; the fibrous serpentine, on the other hand, is associated only with fine- and coarse-grained pentlandite. Contacts between the two textural forms often contain elongate porosity (Fig. 5c, 5f), which ranges in size from 35–240 nm.

Two euhedral magnetite grains (680 and 1000 nm in size), identified by electron diffraction and EDS analysis, occur in close spatial association with the fine-grained pentlandite and are embedded within a matrix of smooth serpentine. SAED patterns of one of the euhedral magnetite grains and a fine-grained pentlandite grain in close proximity show a crystallographic orientation relationship of $[111]_{pn}//[125]_{mgt}$ and $(022)_{pn}//(131)_{mgt}$ (Fig. 5c).

The PS grain composition data are summarized in Table 2 and presented in Figure 3c. We were only able to obtain analyses for the euhedral/subhedral pentlandite along the grain boundaries due to the small size of the rod-like pentlandite. Among all the meteorites, the PS grain pentlandite range from 30.9–33.7 wt. % Ni and from 0.8–1.4 wt. % Co. Each meteorite has a distinct Ni content (MET 01073: 30.9–31.4 wt. %, ALH 84049: 32.5–33.4 wt. %, and ALH 84029: 33.4–33.7 wt. %). The Co compositions of the ALH samples overlap (0.84–1.29 wt. %), but are lower than those of MET 01073 (1.37–1.41 wt. %). Analyses from the ALH pairing group (ALH 84029 and 84049) have higher Ni and lower Co contents than those from MET 01073.

We also performed TEM EDS spot analyses of serpentine in a FIB section extracted from a PS grain (Fig. 5). The data are summarized in the appendix (Table A4). The Fe, Mg, and Si concentrations vary depending on the textural form of serpentine (i.e., smooth versus fibrous). Based on the average of four analyses for each textural group, the smooth serpentine contains more Fe (10.6 ± 0.7 wt. %) and has a lower Mg# (0.77, defined as Mg/(Mg+Fe) in atomic %) than the fibrous serpentine (Fe = 7.2 ± 0.4 wt. % and Mg# = 0.83).

### 3.4 Pyrrhotite+pentlandite+magnetite+serpentine (PPMS) grains

The final group of altered primary sulfide grains is the pyrrhotite+pentlandite+ magnetite+serpentine (PPMS) grains (Fig. 6). These grains share characteristics of both the PPM and the PS grains. They are found in ALH 84029, 84034, and 84049 and are less common than any other textural group. The PPMS grains are only present in the matrix, range in size from 35–125 µm, and are anhedral.

In the PPMS grains, the coarse-grained/patch-textured pentlandite occurs in euhedral to subhedral forms along the periphery of the grains (Fig. 6a–c), whereas the fine-grained pentlandite and blade-textured pentlandite occur in the grain interiors embedded in serpentine (Fig. 6b, 6d). Magnetite, as veins or masses often embedded in serpentine, is limited to the outer portions of the less-altered grains (Fig. 6a), but crosscuts the more-altered grains (Fig. 6c). Similarly, serpentine is present only near the edges in the less-altered grains, but occurs throughout the more-altered grains. Pyrrhotite, when present, occurs in the center of the grains (Fig. 6a–b). In these grains, the textural relationships are suggestive of extensive replacement of pyrrhotite by magnetite, serpentine, and pentlandite with the proportions of the phases to one another varying from grain to grain within the same sample and between different meteorites.

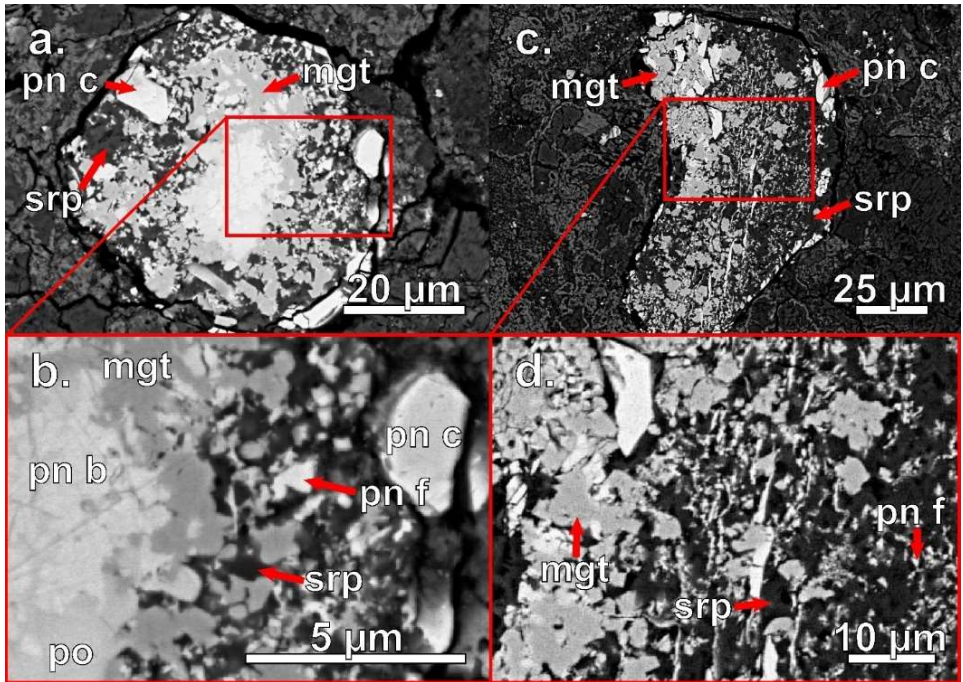

**Figure 6.** BSE images of two PPMS grains from CM1 chondrites show (a) the less-altered PPMS grain containing pyrrhotite near its center, (b) a high magnification image of the grain showing all the observed phases and their relations to one another, (c) the more-altered PPMS grain lacking any observable pyrrhotite, and (d) a high magnification image of the grain showing all the observed phases and their relations to one another. In both grains, coarse-grained pentlandite, surrounded by magnetite, is visible near the grain boundaries. In the less-altered grain (a–b), pyrrhotite is present, and there is less magnetite, serpentine, and fine-grained pentlandite. Po = pyrrhotite, pn = pentlandite, c = coarse grained, f = fine grained, b = blade textured, mgt = magnetite, srp = serpentine.

### 3.5 Other textural types of iron sulfides

In addition to the four textural groups discussed above, there are additional occurrences of sulfides that, in some cases, make up a large proportion of the sulfide in the sample. Examples of these other iron sulfides are included in the appendix (Fig. A3): 1) plain pyrrhotite in MET 01073, characterized by lacking any pentlandite exsolution textures and often having irregular, jagged edges rimmed by serpentine (Fig. A3a–b); 2) coarse-layered sulfides in the ALH pairing group, characterized by ~30–50 μm-sized aggregates of high (sulfide) and low (silicate?) Z phases which form somewhat concentric layers (Fig. A3c–d); and 3) sulfide-rimmed serpentine assemblages in the ALH pairing group, characterized by assemblages (hundreds of μm in size) with serpentine cores rimmed by fine-grained layered sulfides (Fig. A3e–f). These sulfides are, however, texturally distinct from those discussed above and are likely formed by different mechanisms and/or from different precursor materials. We are not considering them further, as the focus of this study is on the alteration of the PPI primary sulfides.

# 4. DISCUSSION

Textural evidence suggests that the 2P, PPM, PS, and PPMS grains are the product of aqueous alteration. However, the different textures and phases among the groups imply that the grains are the result of alteration under different conditions and their mechanisms of alteration were also distinct. As discussed below, the mineralogical evidence strongly supports the view that the observed sulfide alteration in CM1 chondrites is not the result of terrestrial alteration, but is the result of preterrestrial alteration, most likely within an asteroidal environment. We use our mineralogical observations to examine how the different textural groups of altered sulfides formed, what their precursor phases were, and, finally, the constraints that the alteration features place on their formation mechanisms and the conditions of alteration.

## 4.1 Terrestrial versus preterrestrial alteration

We operate on the assumption that these grains were aqueously altered in an asteroidal environment rather than on Earth from weathering processes. The common alteration phases that form on Earth from weathering—goethite ($FeO(OH)$) and maghemite ($Fe_2O_3$) rather than magnetite (Rubin 1997; Cadogan and Devlin 2012; Harju et al. 2014)—do not occur in our altered primary grains. In addition, Lee and Bland (2004) found very little evidence for weathering of troilite in ordinary chondrites recovered from Antarctic, suggesting that weathering of sulfides in the Antarctic environment is very slow. Additionally, these textural groups are present, more or less, in all samples studied but are not present in Antarctica CM2 chondrites. This also suggests that formation of these sulfide alteration textures is not the result of terrestrial weathering processes but is the result of higher degrees of parent body alteration experienced by CM1 chondrites, compared with CM2s.

## 4.2 Textural group occurrences among CM1 samples studied

As summarized in Table 1, the four different sulfide groups occur, with some exceptions, in almost all the chondrites studied. Previous work suggests that the Allan Hills samples (ALH 83100, ALH 84029, 84034, and 84049) are paired (MacPherson 1985a; MacPherson 1985b; Mason 1986). The sulfide mineralogy and textures found in this study support the pairing of ALH 84029, 84034, and 84049, but indicate that ALH 83100 is distinct and not part of this pairing group. However, this does not exclude the possibility that the original bolide was a breccia and that ALH 83100 represents a sample that experienced a different degree of alteration.

The sulfide textural groups that occur in LAP 031166 and MET 01073 differ markedly from the ALH samples and from each other; LAP 031166 and MET 01073 both only contain 2P and PS grains and lack PPM grains. There are clearly similarities between several of the meteorites studied, but some samples are quite distinct. These observations indicate that individual CM1 chondrites did not all experience exactly the same alteration histories.

## 4.3 Formation of CM1 sulfides by alteration of primary pyrrhotite-pentlandite grains

The four textural sulfide groups we have identified have characteristics which suggest that the precursors were similar to the pyrrhotite-pentlandite intergrowth (PPI) grains in CM2 chondrites (Singerling and Brearley 2018). These characteristics include relicts of pyrrhotite and pentlandite with exsolution lamellae as well as the distribution of grains of pentlandite around the periphery of PPI grains. The primary characteristics by textural group that indicate a PPI-like precursor include: pyrrhotite in the PPM and PPMS grains and patch-textured or coarse-grained pentlandite in the PPM, PS, and PPMS grains.

The textural groups in the CM1 sulfides contain other features that are not observed in the CM2 chondrites. These include the presence of serpentine and veins of magnetite and an increase in the proportion of pentlandite in many grains. From these differences, we can conclude that the CM1 sulfide textural groups represent the effects of advanced aqueous alteration of primary sulfides. Sulfide grains in the more-altered CM2 chondrites also exhibit evidence of alteration, with different characteristics; therefore, it is to be expected that in the more extensively altered CM1s, alteration of sulfides should also be more advanced. This conclusion does not imply that CM2 chondrites were the direct precursors of CM1s. Instead, the original, primary PPI grains formed by the same mechanism (i.e., crystallization during chondrule formation) and represent the initial, unaltered primary sulfides present in CM1 and CM2 chondrites.

Additional lines of evidence are consistent with the sulfide textural groups in CM1 chondrites being the product of secondary replacement of primary sulfide grains. For example, magnetite, blade-textured pentlandite, and serpentine occur on the periphery of the grains and extend inward towards the centers, consistent with pseudomorphic replacement. Interfaces between these phases and those that occur in the PPI grains are often irregular and jagged. Additionally, the secondary characteristics by textural group include: phyllosilicate lenses and the dominance of pentlandite in the 2P grains; magnetite and blade-textured pentlandite in the PPM grains; serpentine and fine-grained pentlandite in the PS grains; and magnetite, serpentine, and blade-textured and fine-grained pentlandite in the PPMS grains. Finally, these textures are not observed in the least-altered CM2 chondrites.

In the following discussion, we make a detailed comparison of the style and degree of alteration of sulfides in the CM2 and CM1 chondrites. We pull heavily from our observations of altered sulfides in CM2 chondrites in Singerling and Brearley (2020). By performing this comparison, we hope to determine if the two petrologic types can be interpreted as the result of progressive alteration, with CM2s representing either an earlier stage of alteration or the products of alteration of similar starting materials but not otherwise related to one another (i.e., derived from different parent bodies; e.g., Rubin et al. 2007; Lee et al. 2019).

*4.3.1 Porous pentlandite (2P) grains*

A comparison of the 2P grains in CM1 chondrites to the porous pyrrhotite-pentlandite (3P) grains in CM2 chondrites (Singerling and Brearley 2020) suggests that although the two textural groups have similarities, they are not genetically related. In other words, the 3P grains in the CM2 chondrites did not alter to form the 2P grains in the CM1 chondrites. Instead, both the 2P and 3P grains likely represent the products of alteration, starting with similar precursor materials but altering under differing conditions and to different extents.

The increase in abundance of pentlandite from the 3P to the 2P grains would be consistent with progressive replacement of pyrrhotite by secondary pentlandite, but there are distinct differences that make the argument for a genetic-link between the two unsatisfactory. . . The least-altered 3P grains (e.g., CM2 QUE 97990 from Singerling and Brearley 2020) contain more pyrrhotite than pentlandite, whereas 3P grains in more-altered CM2s (e.g., CM2 Mighei from Singerling and Brearley 2020) contain more pentlandite with only minor pyrrhotite. The observations from the CM2 3P grains imply that nearly all pyrrhotite is altered to pentlandite even within only the more-altered 3P grains. If the 3P and 2P grains are genetically-linked, we would expect little pyrrhotite to remain even prior to the alteration stages seen in CM1. However, the CM1 2P grain in CM1 ALH 84049 shows obvious pyrrhotite relicts in the form of twinning (Fig. 2d–f). Additionally, this twinning was not observed in the CM2 3P grains.

In the least-altered 3P grains, pores primarily occur in the pyrrhotite either perpendicular or parallel to the pentlandite lamellae. In the more-altered 3P grains, pores occur throughout the grains with no apparent preference for pentlandite or pyrrhotite and are primarily oriented parallel to the remnant pyrrhotite slivers. In the 2P grains, pore-like features are filled with phyllosilicates and likely represent pores formed from dissolution of pentlandite that were subsequently filled with Si,Mg-bearing fluids which precipitated phyllosilicates. Taken together, these differences imply that the 2P and 3P grains are not genetically linked and represent alteration along different paths.

*4.3.2 Pyrrhotite+pentlandite+magnetite (PPM) grains*

The PPM grains in CM1 chondrites have textural similarities to altered PPI grains in CM2 chondrites (Singerling and Brearley 2020), which indicates that the textural groups had similar precursors. However, the differences in textures between the two and the presence of blade-textured pentlandite in the CM1 PPM grains indicate distinct alteration paths for the two petrologic types.

The textural characteristics of the CM2 altered PPI grains imply that they represent relicts of primary sulfides. These include the presence of coarse-grained pentlandite along the periphery of the grains as well as fine-grained pentlandite lamellae, blades, and/or rods embedded in pyrrhotite in the interior of the grains. These textures are also apparent in the CM1 PPM grains suggesting that these grains also represent surviving primary sulfide relicts. The altered PPI grains in CM1 and CM2 chondrites contain both primary pyrrhotite and pentlandite.

In one subset of the CM2 altered PPI grains, referred to as PPI alt mgt grains hereafter (e.g., Singerling and Brearley 2020), the pyrrhotite has been replaced by patches of porous magnetite. The boundaries between the magnetite and pyrrhotite have flame-like textures extending from the magnetite into the pyrrhotite. In the CM1 PPM grains, pyrrhotite has been replaced by magnetite veins, which are not porous even on the TEM scale (Fig. 4c–e). The magnetite appears to be composed of numerous interlocking subgrains which do not display a flame-like texture when in contact with pyrrhotite.

In the CM2 PPI alt mgt grains, the only pentlandite observed has textures consistent with primary pentlandite and appears to be largely resistant to replacement by the magnetite with relict pentlandite visible as patches on the edge of the grain and as lamellae. In the CM1 PPM grains, on the other hand, both relict pentlandite, present as patches near grain peripheries, in addition to a blade-textured pentlandite, often occurring between the magnetite veining and pyrrhotite (Fig. 4c), were observed.

While the phases (pyrrhotite, pentlandite, and magnetite) are similar between the CM2s and the CM1s, the textures are distinctly different, especially regarding magnetite (porous patches in the CM2s versus non-porous veins in the CM1s). Additionally, the CM2 PPI alt mgt grains do not contain a secondary (i.e., blade-textured) pentlandite. The presence of relict pyrrhotite with the characteristics of primary sulfide indicates that they had the same precursor, but the textural characteristics of the magnetite and the formation of blade-textured pentlandite in the CM1 PPM grains indicates that the alteration pathways and degree of alteration were different between the CM2 and CM1 chondrites. This implies that the CM1 PPM grains are not examples of more advanced alteration of the CM2 PPI alt mgt grains, but rather, the two experienced different alteration paths, as discussed later.

*4.3.3 Pentlandite+serpentine (PS) grains*

An important textural feature of the CM1 PS grains is the presence of coarse-grained pentlandite located on the periphery of the grains, a feature that is diagnostic of primary sulfides in CM2 chondrites (Singerling and Brearley 2020). In the CM1 PS grains, no primary pyrrhotite remains, although the overall morphology of the primary PPI grain, with the pyrrhotite originally in the interior of the grains and the coarse-grained pentlandite on the rim, is still intact.

However, the differences in textures between the two and the presence of fine-grained secondary pentlandite in the CM1 PS grains indicate different alteration paths that suggest very distinct alteration histories for the two petrologic types. In one subset of the CM2 altered PPI grains, referred to as PPI alt phy grains hereafter (e.g., Singerling and Brearley 2020), the pyrrhotite has been replaced by patches of phyllosilicates. The proportion of phyllosilicates varies by grain, but rarely shows complete replacement of pyrrhotite even in the more-altered CM2 chondrites. In contrast, the CM1 PS grains are dominated by serpentine that completely replaces pyrrhotite (Fig. 5a). Additionally, the phyllosilicates in the CM2 PPI alt phy grains appear layered and fibrous, while the serpentine in the CM1 PS grains varies from a smooth to a fibrous, unlayered texture.

In the CM2 PPI alt phy grains, pentlandite, occurring as patches primarily on the periphery of the grains, has been largely resistant to replacement by phyllosilicates. Pentlandite in the CM1 PS grains, on the other hand, is of two types: 1) primary, coarse-grained pentlandite along the periphery of grains similar to the CM2 PPI alt phy grains and 2) secondary, fine-grained pentlandite in the interior of the grains embedded in serpentine.

TEM observations reported by Brearley (2011) on PPI alt phy grains in the CM2 chondrite Mighei can be compared to our TEM work on the CM1 PS grains. Brearley (2011) found the presence of a fibrous oxysulfide, in addition to an iron oxide (likely magnetite) as alteration products. Our TEM studies of the PS grains show the alteration products include serpentine, pentlandite, and minor amounts of magnetite. In conclusion, the PS grains seem to have had PPI grains as their precursor, but like the PPM grains, they do not represent more advanced alteration of PPI alt grains from CM2 chondrites, but instead followed different alteration pathways as a result of differing alteration conditions.

*4.3.4 Compositions*

We compare the compositions of sulfides in CM2 and CM1 chondrites in Figure 7 to better understand how the two sets of sulfides are related. Figure 7a compares the compositions of individual sulfide phases in PPI, PPM, and PS grains, whereas Figure 7b compares the bulk compositions of 2P and 3P sulfide grains. The CM2 analyses come from Singerling and Brearley (2018) for the unaltered primary grains (PPI) and Singerling and Brearley (2020) for the 3P grains. To summarize from our previous discussions: primary sulfides include all pyrrhotite, CM2 PPI pentlandite, CM1 PPM patch-textured pentlandite, and CM1 PS coarse-grained pentlandite; secondary phases include CM2 3P pentlandite, CM1 2P pentlandite, and CM1 PPM blade-textured pentlandite.

A comparison of CM2 PPI grains (i.e., primary sulfide compositions) to the CM1 PPM and PS grains (Fig. 7a) illustrates several differences. The primary CM2 pyrrhotite has a larger range in Co and Ni contents, which overlaps with the CM1 pyrrhotite compositions. However, the CM2 pyrrhotite tends to plot along a trend that is more Co rich and Ni poor than the CM1 pyrrhotite. The primary CM2 pentlandite overlaps with the primary CM1 pentlandite (i.e., PPM patch-textured and PS coarse-grained pentlandite), especially in Co content, consistent with our hypothesis that they are indeed primary pentlandites. The CM1 pentlandite typically has higher Ni contents, with a smaller compositional range in Ni and a larger range in Co. The secondary CM1 pentlandite (i.e., PPM blade-textured pentlandite), on the other hand, plots at lower Co compositions (<0.6 wt. %). The compositional, as well as textural, differences between the blade-textured and patch-textured pentlandite in the PPM grains imply that the blade-textured pentlandite is a secondary phase, consistent with this hypothesis. Comparing the pentlandite from the PPM and PS grains to the 2P and 3P grains (envelopes in Fig. 7a), we find that the primary pentlandite overlaps with the 3P grains and the lower Co 2P grains. However, the blade-textured pentlandite is notably lower in Co content.

   Comparing individual analyses from CM2 3P and CM1 2P grains (envelopes in Fig. 7a), several differences are apparent. The CM2 3P grains have a larger range in Ni content, but have compositions which overlap with the range for CM1 2P grains. The Co contents differ significantly between the two petrologic types. Comparing the bulk compositions of the CM2 3P and CM1 2P grains (Fig. 7b), a clear divide occurs between the 3P and 2P grain Co contents, consistent with there not being a genetic link between the two. Although a small number of analyses from the 2P grains have Co contents similar to analyses from the 3P grains (0.92−1.56 wt. %), the majority of the analyses plot at higher Co contents (>2 wt. %). Additionally, from the bulk compositions, we find that with increasing alteration of the bulk meteorite, the 3P/2P grain compositions follow a trend of increasing Co and, to a lesser extent, Ni contents. This trend is apparent in the CM2 chondrites, which, based on the mineralogic alteration index of Browning et al. (1996) and the petrologic sequence of Rubin et al. (2007), increase in degree of alteration from QUE 97990 (2.6) to Murchison (2.5) to Murray (2.5) to Mighei (not studied by Rubin et al., 2007 but listed as more altered than Murray in Browning et al., 1996). See Singerling and Brearley (2020) for more detail on these petrologic sequences. Although they are not the product of the alteration of the 3P grains, the 2P grains extend this general trend of higher Co contents. The 2P grains do not, however, show an increase in Ni contents given that they overlap with the values of the 3P grains. In summary, we observe that: 1) PPM grain blade-textured pentlandite compositions have lower Co contents but similar Ni contents compared with PPI grain primary pentlandite; 2) 3P grain pentlandite compositions have slightly higher Co contents but similar Ni contents compared with PPI grain primary pentlandite; and 3) 2P grain pentlandite compositions have higher Co contents and slightly higher Ni contents compared with 3P grain pentlandite. These trends of changing Co and Ni contents are consistent with textural observations arguing that the blade-textured pentlandite and the pentlandite in the 3P and 2P grains are secondary in origin. However, we observe opposing behaviors. In the PPM grains, secondary pentlandite (blade textured) has lower Co contents compared to primary pentlandite (patch textured), but in the 2P grains we observe the highest Co contents among any sulfide phase in the CM2 and CM1 samples.

   Previous studies of the Fe-Ni-S system have demonstrated that the $fO_2$ of the system affects the Co content of pentlandite with higher $fO_2$ correlating to higher Co contents (Schrader et al. 2016). This would be consistent with the 2P grains having undergone alteration under higher $fO_2$ than the PPM grains; however, the presence of magnetite in the latter implies that the PPM grains experienced alteration under oxidizing conditions. Schrader et al (2021) argue that the Fe/S ratio of low-Ni pyrrhotite can be used as a proxy for the extent of oxidation experienced by the sulfide, with lower values corresponding to more oxidizing conditions. As Figure 7c shows, the Fe/S and cation/S, defined as (Fe+Co+Ni+Cr)/S, ratios of the pyrrhotite in the PPM grains is much lower than that in unaltered CM2 pyrrhotite (PPI) and altered CM2 pyrrhotite (PPI alt mgt). The CM2 pyrrhotite cation/S ratios overlap and range from 0.96–1.03, whereas the CM1 PPM pyrrhotite cation/S ratios are significantly lower, ranging from 0.90–0.92. The complete data are available in the appendix (Table A5). If the PPM grains then experienced alteration under more oxidizing conditions, we would expect the Co content to be higher in the secondary pentlandite (blade-textured) in the PPM grains; however, we observe the opposite behavior. Instead, the original Co content of the primary pyrrhotite may have also played a role in addition to $fO_2$. In the PPM grains, the relict pyrrhotite has low Co contents consistent with the compositions of the precursor PPI grains; Co is preferentially incorporated into pentlandite during pyrrhotite-pentlandite exsolution from monosulfide solid solution (*mss*) (e.g., Soltanieh et al. 1990; Raghavan 2004; Dare et al. 2010). *Mss* ((Fe,Ni)$_{1-x}$S) is the stable sulfide phase at temperature >870 K (Kitakaze et al. 2011).

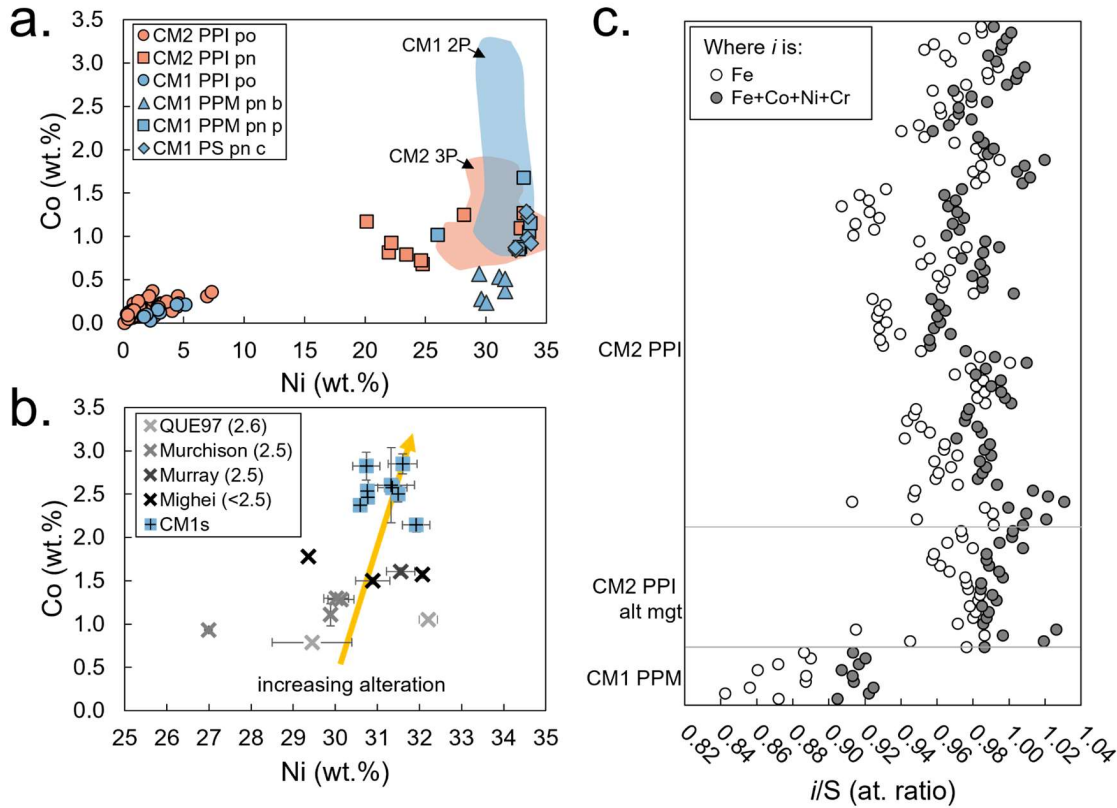

**Figure 7.** Elemental composition plots comparing sulfide phases in CM2 and CM1 chondrites. (a–b) show Co versus Ni plots (in wt. %) of (a) individual spot analyses from pyrrhotite or pentlandite in the PPI (CM2s), PPM (CM1), and PS (CM1) grains and (b) bulk compositions for the 3P (CM2) and 2P (CM1) grains calculated as an average for a grain, based on several spot analyses, with error bars of 1σ. In (a) envelopes for the spot analyses from 3P (CM2) and 2P (CM1) grains are included for comparison purposes. In (b), CM2 3P grains are grouped by sample, with the number in the legend reflecting the extent of alteration of the bulk sample, using the alteration scheme in Rubin et al. (2007). CM2 sample symbols increase in darkness with increasing alteration of the overall meteorite in which they occur. (c) shows Fe/S and cation/S ($i$/S) atomic ratios of pyrrhotite from CM2 unaltered (PPI), CM2 altered (PPI alt mgt), and CM1 (PPM) grains. The cation/S ratio, defined as (Fe+Co+Ni+Cr)/S, more accurately reflects the stoichiometry of the pyrrhotite, since the cations listed can substitute for Fe in the pyrrhotite structure. CM2 data are from Singerling and Brearley (2018, 2020). Po = pyrrhotite, pn = pentlandite, p = patch textured, b = blade textured, c = coarse grained.

We propose that when pyrrhotite in the PPM grains altered to pentlandite, the Co content of the secondary pentlandite was inherited directly from the pyrrhotite and hence has lower Co contents than the primary pentlandite in the same grains. In the case of the 2P grains, $fO_2$ rather than the Co content of the precursor pyrrhotite may have had a stronger effect on the Co content of the secondary pentlandite. In either case, different processes, themselves a function of alteration mechanisms and/or conditions, controlled the Co contents of the different secondary pentlandites. In the following section, we explore differences in alteration mechanisms and conditions which might explain the different compositions of the secondary pentlandite phases as well as the presence of the other secondary phases observed.

**4.4 Alteration mechanisms and conditions**

As mentioned previously, we propose that the 2P, PPM, PS, and PPMS grains in CM1 chondrites were all initially PPI grains, which experienced alteration under different conditions resulting in different textural groups. We now discuss the possible mechanisms and conditions that resulted in the formation of these altered sulfides and how these compare to the altered sulfides in CM2 chondrites.

*4.4.1 Porous pentlandite (2P) grains*

We propose that the 2P grains were initially PPI grains which may have had an intermediate stage similar to, but distinct from the 3P grains. The 2P grains are characterized by phyllosilicate lenses and the absence of pyrrhotite, implying that the grains experienced dissolution followed by precipitation of phyllosilicates as well as replacement of pyrrhotite by pentlandite. The similarities between the 3P and 2P grains argues for a similar formational history. That does not imply that the CM1 and CM2 grains are genetically linked, but rather had similar formation mechanisms. Figure 8a illustrates a schematic representation of the stages of alteration.

The alteration would have been instigated by a change in conditions that promoted dissolution of the PPI grain. This could have been the result of changes in several different variables, including a decrease in the pH, an increase in the $fO_2$, and/or changes in the fluid composition, represented by the activities of certain components in solution. Figure 9a is a log $fO_2$-pH diagram calculated using Geochemist Workbench® showing the different possible scenarios in which pyrrhotite would become unstable and begin to experience dissolution (represented by the $Fe^{2+}$ stability field). A decrease in the pH would cause pyrrhotite (Point A1) to become unstable and move into the $Fe^{2+}$(aq) stability field (Point A2). An increase in $fO_2$ would also cause the pyrrhotite (Point B1) to become unstable and move in the $Fe^{2+}$(aq) stability field (Point B2). Lastly, a change in fluid composition (e.g., decrease in iron activity from $10^{-6}$ to $10^{-8}$) would cause the pyrrhotite (Point C) to become unstable, plotting in the $Fe^{2+}$(aq) stability field. According to this diagram, it is not possible to achieve dissolution with a change in temperature for the range of temperatures expected during aqueous alteration for most of the CM chondrite parent body (i.e., <100°C; DuFresne and Anders 1962; Clayton and Mayeda 1984; Guo and Eiler 2007).

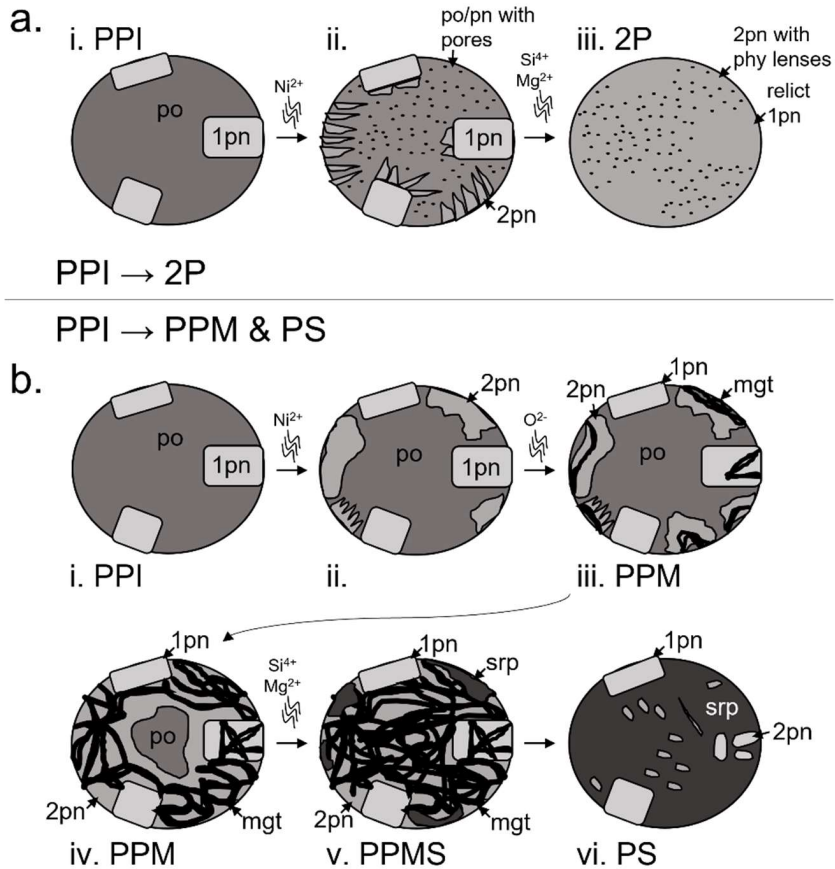

**Figure 8.** Schematic diagrams representing the stages of alteration of PPI grains to form: (a) 2P grains—(i) PPI grain is stable prior to alteration having originally formed in the solar nebula from crystallization of sulfide melts during chondrule formation. (ii) Changes in environmental conditions (pH, $fO_2$, Fe activity, fluid composition, etc.) cause dissolution of pyrrhotite, forming numerous submicron pores, and transformation into secondary pentlandite. (iii) Continued alteration causes all primary pyrrhotite to be replaced by secondary pentlandite, and the introduction of Si,Mg-bearing fluid causes precipitation of phyllosilicates in pore space, resulting in 2P grains—and (b) PPM and PS grains—(i) PPI grain is stable prior to alteration. (ii) The introduction of Ni-bearing fluid causes pentlandite to replace pyrrhotite. (iii) Changes in conditions ($fO_2$) cause magnetite to form from secondary pentlandite, resulting in PPM grains. (iv) Continued alteration causes more pyrrhotite to be replaced by secondary pentlandite and that in turn by magnetite. (v) Additional changes in environmental conditions (fluid composition) cause replacement of magnetite by serpentine, resulting in PPMS grains. (vi) Continued alteration causes more magnetite to be replaced by serpentine until only primary and secondary pentlandite and serpentine remain, resulting in PS grains. Po = pyrrhotite, 1pn = primary pentlandite, 2pn = secondary pentlandite, mgt = magnetite, srp = serpentine.

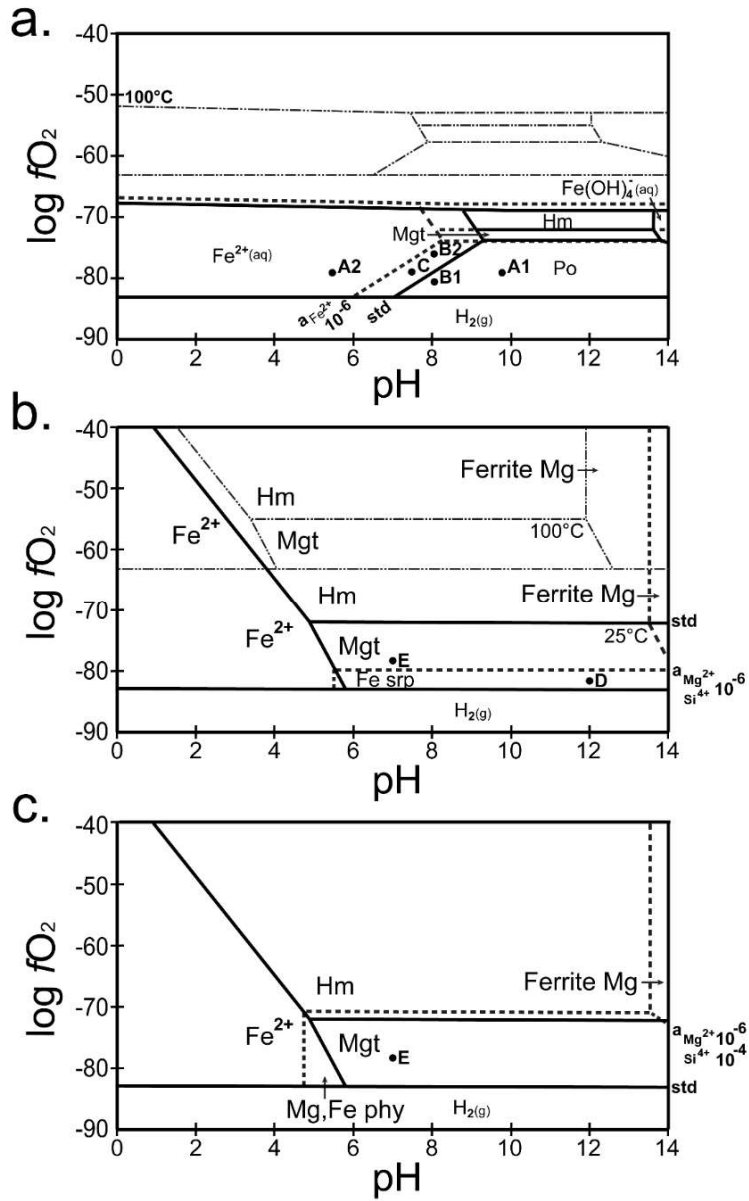

**Figure 9.** Log $fO_2$ versus pH diagrams, calculated using the Geochemist Workbench®, for (a) pyrrhotite interacting with $Fe^{2+}$-bearing aqueous fluid and (b–c) magnetite interacting with $Mg^{2+}$- and $Si^{4+}$-bearing aqueous fluid. Solid lines show standard conditions (std): T = 25°C, P = 1 bar, $aH_2O = 1$, and $a_i = 10^{-8}$, where i = $Fe^{2+}$ for (a) and $Mg^{2+}$ and $Si^{4+}$ for (b–c). Thin dashed lines show changes in T to 100°C in (a) and (b). Thick dashed lines show changes in $a_i$ in aqueous solution as follows: (a) increase in $aFe^{2+}$ to $10^{-6}$, (b) increase in $aMg^{2+}$ to $10^{-6}$ and $aSi^{4+}$ to $10^{-6}$, and (c) increase in $aMg^{2+}$ to $10^{-6}$ and $aSi^{4+}$ to $10^{-4}$. In (a), an increase in $aFe^{2+}$ expands the solid stability fields to lower pH and sees the $Fe(OH)_4^-$ (aq) field disappear. In (b–c), an increase in $aMg^{2+}$ and $aSi^{4+}$ changes the stability field of magnetite to serpentine (b) or phyllosilicates (c). In general, an increase in T shifts the stability of fields to higher $fO_2$ and lower pH values. Points A1−E are discussed in the text. Po = pyrrhotite, mgt = magnetite, hm = hematite, srp = serpentine, phy = phyllosilicates.

As the 2P grains show, the dissolution of pyrrhotite was also associated with replacement of pyrrhotite by pentlandite. Figure 9a does not show pentlandite due to the limitations inherent in the Geochemist Workbench®; pentlandite is not included in the software database and so is not represented in any diagrams generated. A possible reaction, in two simplified forms and then a more stoichiometrically accurate form, is:

$$pyrrhotite + Ni^{2+}(aq) \rightarrow pentlandite + Fe^{2+}(aq)$$
$$Fe_7S_8(s) + Ni^{2+}(aq) \rightarrow Fe_{4.5}Ni_{4.5}S_8(s) + Fe^{2+}(aq)$$
$$2Fe_5^{2+}Fe_2^{3+}S_8(s) + 9Ni^{2+}(aq) + 2H_2O(l) \rightarrow 2Fe_{4.5}^{2+}Ni_{4.5}^{2+}S_8(s) + 5Fe^{2+}(aq) + O_2(aq) + 4H^+(aq)$$

The above equations illustrate that in order to form pentlandite from pyrrhotite, the pyrrhotite must gain $Ni^{2+}$ and lose $Fe^{2+}$. Additionally, the insolubility of $Fe^{3+}$ for the range of pHs expected for CM chondrites requires the reduction of the $Fe^{3+}$ in pyrrhotite to $Fe^{2+}$ (e.g., Janzen et al. 2000; Belzile et al. 2004). The reaction also shows a decrease in the pH moving from products to reactants with the generation of $H^+$.

The $Ni^{2+}$ (aq) in the above equations could have initially come from the breakdown of Fe,Ni metal during aqueous alteration since metal is observed to be one of the earliest phases to alter in CM2 chondrites (e.g., Tomeoka and Buseck 1988; Hanowski and Brearley 2001; Rubin et al. 2007). However, due to the highly altered state of CM1s, Fe,Ni metal may have already broken down at earlier stages of alteration. An alternate source of Ni could have been from the breakdown of tochilinite, which contains ~5±2 wt. % Ni in CM2 chondrites (Palmer and Lauretta 2011). Tochilinite is stable under reducing conditions and at temperatures below 120°C (Browning and Bourcier 1996; Zolensky et al. 1997) and is a common phase in CM2 chondrites. However, tochilinite is not observed in the CM1 chondrites, so it either formed at an earlier CM2-like stage and completely broke down or never formed at all.

Assuming the former, we performed a simple mass balance calculation to determine if Ni-bearing tochilinite in CM chondrites could be the potential source of Ni to explain the observed abundances of secondary pentlandite in CM1 chondrites. Details of the calculation are reported in Appendix A. The calculated minimum and maximum abundance values, based on the variable Ni contents of tochilinite, pyrrhotite, and pentlandite, yield 0.1 and 1.7 vol. % pentlandite, respectively. These values are comparable to the abundances of pentlandite in CM1 chondrites (0–1.2 vol. % from 4 CM1 chondrites in Howard et al. 2011), implying that tochilinite is a realistic potential source of the Ni required for the formation of secondary pentlandite. It is also possible that CM1 and CM2 chondrites are not related, and tochilinite never formed on the CM1 parent body to begin with. In that case, an alternative source of Ni is required, perhaps from the initial breakdown of Fe,Ni metal at much earlier stages of alteration, especially if that Ni then remained in solution rather than forming tochilinite.

The presence of phyllosilicate lenses in the 2P grains is another secondary feature, one not observed in the 3P grains. We argue that these lenses were originally pore space that was filled with phyllosilicates that precipitated from Si,Mg-bearing fluids. This introduction of fluid likely occurred at a later stage than the Ni-bearing fluid that formed the secondary pentlandite. Pentlandite is more resistant to alteration by Si,Mg-bearing fluids compared to pyrrhotite, as evidenced by the PPI alt phy grains in Singerling and Brearley (2020), which is likely why the phyllosilicates are limited to formation within pore space.

In summary, the 2P grains are the products of advanced alteration of the PPI grains by dissolution and replacement of pyrrhotite by secondary pentlandite followed by precipitation of phyllosilicates in pore space. This reaction requires the addition of $Ni^{2+}$ and a later introduction of Si,Mg-bearing fluids. The alteration of the PPI grains to the 2P grains in CM1 chondrites provides evidence for acidic and/or oxidizing conditions and changing fluid compositions, specifically Ni-bearing and later Si,Mg-bearing fluids.

*4.4.2 Pyrrhotite-pentlandite-magnetite-serpentine (PPM, PS, and PPMS) grains*

The PPM, PS, and PPMS grains are all characterized by remnant pentlandite exsolution textures with replacement of pyrrhotite by pentlandite, magnetite, and/or serpentine. The PPMS grains represent a transition between the PPM and PS grains implying that the two are genetically related. We propose that the PPM, PS, and PPMS grains were all initially PPI grains, which altered into the PPM grains and then, with further alteration, into the PPMS grains and finally into the PS grains. The occurrence of different sulfide textural groups within meteorite samples which have experienced different degrees of aqueous alteration further supports this hypothesis. MET 01073 contains textural features distinct from the ALH samples. We can postulate that the sulfide textural groups in this CM1 sample reflect more extensive alteration. Tellingly, MET 01073 contains PS and 2P grains but does not contain PPM grains. From their occurrence within the meteorite samples, we hypothesize that the PPM grains formed from moderate alteration, while the PS grains are the result of more extensive alteration.

Figure 8b is a schematic representation of the stages of alteration. Similar to the 3P/2P grains, the first stage involved pyrrhotite altering to secondary pentlandite (e.g., blade-textured pentlandite) via the same reaction described above. In the case of the PPM grains, however, porosity did not develop in this pentlandite. Additionally, magnetite veining was observed in all PPM grains, usually within pentlandite but less commonly within pyrrhotite. Following the formation of the secondary pentlandite, the PPM grains experienced a change in conditions that promoted the oxidation of the secondary pentlandite to form magnetite. A possible reaction, in two simplified forms and then a more stoichiometrically accurate form, for the formation of the PPM grain magnetite from secondary pentlandite is:

$$pentlandite + O_2(aq) \rightarrow magnetite + Ni^{2+}(aq) + S^{2-}(aq)$$
$$Fe_{4.5}Ni_{4.5}S_8(s) + O_2(aq) \rightarrow Fe_3O_4(s) + Ni^{2+}(aq) + S^{2-}(aq)$$
$$2Fe_{4.5}^{2+}Ni_{4.5}^{2+}S_8(s) + \frac{15}{2}O_2(aq) + 6H^+(aq) \rightarrow 3Fe^{2+}Fe_2^{3+}O_4(s) + 9Ni^{2+}(aq) + 16S^{2-}(aq) + H_2O(l)$$

The above equations illustrate that in order to form magnetite from pentlandite, the pentlandite must gain $O_2$ and lose $Ni^{2+}$ and $S^{2-}$. Hence, the system must become oxidizing. Additionally, the insolubility of $Fe^{3+}$ limits the availability of this species in solution. Instead, the $Fe^{3+}$ required for magnetite formation was likely sourced directly from $Fe^{2+}$ in the pentlandite and was oxidized during the reaction. The reaction also shows an increase in the pH moving from products to reactants as consumption of $H^+$ occurs.

As noted previously, we propose that the PPM grains represent an intermediate stage in the alteration of primary sulfides that evolve with more extensive alteration into the PS grains. This hypothesis is supported by relative abundance of the different textural groups of altered sulfides in meteorites with varying degrees of alteration. The less-altered CM1 chondrites have a higher abundance of PPM grains compared to PS grains, while the more-altered CM1 chondrite (MET 01073) does not contain PPM grains, but does contain PS grains. Additionally, the PPM grains contain primary pyrrhotite, while it is notably absent in the PS grains. Again, the presence of the PPMS grains implies that the PPM and PS grains are genetically related to one another. Therefore, we can assume that the PS grains represent the products of the continued alteration of the PPM grains.

Such a scenario, presented in Figure 8b, involved the complete replacement of magnetite with serpentine in a sequence of steps. First, primary pyrrhotite was replaced by secondary pentlandite, which is itself replaced by magnetite, apparent in some PPM grains (Fig. 4b–c). Then, with changes in conditions, such as the fluid composition, magnetite became unstable and reacted with ions in solution, specifically $Mg^{2+}$ and $Si^{4+}$, to form serpentine. The $Mg^{2+}$(aq) and $Si^{4+}$ (aq) likely originally came from the breakdown of forsteritic olivine, one of the last phases to alter in CM chondrites (e.g., Hanowski and Brearley 2001; Howard et al. 2011). The final assemblage in these altered sulfide grains is primary pentlandite, secondary pentlandite (as formed in the PPM grains), and serpentine.

The alteration of the PPM grains into the PS grains would have been initiated by a change in fluid composition as illustrated in Figure 9b–c. For example, an increase in the activities of $Mg^{2+}$ and $Si^{4+}$ in the fluid from $10^{-8}$ to $10^{-6}$ (Fig. 9b), driven by dissolution of forsteritic olivine, causes the stability field of magnetite to contract at lower $fO_2$ values, and the field of Fe serpentine expands. Depending on the $fO_2$ of the system, the magnetite could become unstable under these fluid compositions (Point D) and alter into a serpentine, although there is a range of $fO_2$ values where the magnetite is still stable (Point E). Increasing the activity of $Si^{4+}$ in the fluid from $10^{-6}$ to $10^{-4}$ (Fig. 9c) causes the stability field of magnetite to be completely replaced by a Mg,Fe phyllosilicate. There are fluid compositions where magnetite is unstable regardless of $fO_2$ and alters into phyllosilicates/serpentines (Point E). High temperatures (i.e., 100°C) suppress the formation of serpentine or phyllosilicates (Fig. 9b).

A possible reaction for the formation of PS grains from PPM grains is presented below in only simplified terms due to the uncertainty in the chemical formula, and more specifically the valence state of iron, for the serpentine:

$$magnetite + Si^{4+}(aq) + Mg^{2+}(aq) + H_2O(l) \rightarrow serpentine$$
$$Fe_3O_4(s) + Si^{4+}(aq) + Mg^{2+}(aq) + H_2O(l) \rightarrow (Fe,Mg)_3Si_2O_5(OH)_4(s)$$
$$2Fe^{2+}Fe_2^{3+}O_4(s) + 2Si^{4+}(aq) + Mg^{2+}(aq) + 3H_2O(l) \rightarrow Fe_2^{2+}Mg^{2+}Si_2O_5(OH)_4(s) + 3Fe^{2+}(aq) + 2H^+(aq) + O_2(aq)$$

Terrestrial analogues for the PPM and PS grains have been observed in serpentinized layered mafic intrusions, such as the UG2 and Merensky Reef of the Bushveld Complex in South Africa (Li et al. 2004), and serpentinized ultramafic cumulate bodies, such as the Black Swan disseminated orebody in Western Australia (Barnes et al. 2009). In the former case, pyrrhotite, pentlandite, and chalcopyrite have been replaced by magnetite or serpentine. In the latter case, pyrrhotite has been replaced by magnetite, ferroan brucite, and ferroan magnesite via oxidation and carbonation reactions. Barnes et al. (2009) argue that the reaction of pyrrhotite to magnetite, brucite, and magnesite yielded Ni-rich sulfides (i.e., pentlandite, heazlewoodite, millerite) in a subset of the sulfide assemblages in the Black Swan orebody. In some PPM grains, we observe magnetite in direct contact with pyrrhotite, so it is plausible that magnetite can form directly from pyrrhotite. In either case, the terrestrial sulfide alteration assemblages are consistent with the predominance of pentlandite rather than pyrrhotite in the CM chondrite PS grains.

The characteristics of alteration of iron sulfide to phyllosilicates is distinct in the CM2 (PPI alt phy grains) and CM1 (PS grains) chondrites. Different phases are present (pyrrhotite, pentlandite, and phyllosilicate in PPI alt phy grains; pentlandite and serpentine in PS grains), different textural features (coarse-grained pentlandite and a fibrous phyllosilicate in PPI alt phy grains; coarse- and fine-grained pentlandite and smooth and fibrous serpentine in PS grains), and different degrees of replacement of pyrrhotite (partial in PPI alt phy grains; complete in PS grains). These differences can be explained by the formation of PPI alt phy grains by direct replacement of pyrrhotite by phyllosilicates, while the PS grains formed in a sequence of reactions where pyrrhotite was first replaced by pentlandite which was replaced by magnetite which was replaced by serpentine. These distinct alteration mechanisms are further supported by the lack of sulfur or nickel in the serpentine of the PS grains and the presence of sulfur in the phyllosilicates of the PPI alt phy grains, reported by Brearley (2011).

*4.4.3 Trends in modal abundances with different degrees of aqueous alteration*

Clearly, there are similarities and differences between the alteration mechanisms and conditions of the four textural groups of sulfides. Overall, we observe that primary pyrrhotite is unstable under specific alteration conditions and breaks down into 1) secondary pentlandite in the 2P grains, 2) secondary pentlandite then magnetite in the PPM grains, and 3) serpentine in the PS grains from the magnetite in the PPM grains. On the contrary, primary pentlandite is largely stable and resistant to alteration. Therefore, compared to the CM2 chondrites, in the CM1 chondrites we would expect to see a decrease in the abundance of pyrrhotite and an increase in pentlandite, magnetite, and serpentine, resulting from more extensive alteration.

Figure 10 illustrates the relative abundances of the different sulfide grain types in the CM2 chondrites, from Singerling and Brearley (2020), and CM1 chondrites studied in this work. With increasing alteration, there is a decrease in PPI grains and an increase in the altered sulfides: 3P, PPI alt mgt, PPI alt phy, 2P, PPM, PPMS, and PS grains. Note that we have not measured the absolute modal abundance of these phases, but are basing our assessment on the relative occurrence of the different types of sulfide grains. With that in mind, a decrease of PPI grains corresponds to a decrease of pyrrhotite and an increase in pentlandite, magnetite, and serpentine. The expected increase in the modal abundance of magnetite would correlate with an increase in the iron oxidation state of the bulk meteorite. This is consistent with the bulk Fe oxidation states ($Fe^{3+}/(Fe^{2+}+Fe^{3+})$) measured by bulk XANES techniques (Beck et al. 2012), which increase from the CM2 to the CM1 chondrites.

The expected changes in modal abundances of pyrrhotite, pentlandite, magnetite, and serpentine are largely consistent with studies of mineralogical modal abundances in bulk CM2 and CM1 chondrites. From CM2 to CM1 chondrites, the modal abundances (in average vol. % from 24 meteorites) change as follows: sulfides increase from 1.8% to 2.0%, magnetite increases from 1.5% to 2.4%, and Mg-rich serpentine increases from 38.9% to 61.9% (data from: Howard et al. 2011; King et al. 2017). The iron sulfides are further broken down into pyrrhotite and pentlandite from Howard et al. (2009, 2011) (data from the 2011 paper is derived from a figure), but their findings show that, from CM2 to CM1 chondrites, pyrrhotite increases from 1.4% to 1.8% and pentlandite decreases from 0.4% to 0.3%. On the other hand, Villalon et al. (2021) showed that the matrix nanosulfide population increases in the abundance of pentlandite over pyrrhotite in more highly-altered lithologies of CM2 Paris.

This disagreement between our observations and the modal abundance of sulfides can be explained by the fact that we selected a subset of grains for study; we did not take into account the presence of additional iron sulfide textural groups, as discussed previously in the Results section. The modal abundances of pyrrhotite and pentlandite also include the grains of pure pyrrhotite in MET 01073 and the coarse-layered sulfides and sulfide-rimmed serpentine assemblages in the ALH pairing group, as well as any other sulfide grain types that we did not use as the basis for this work.

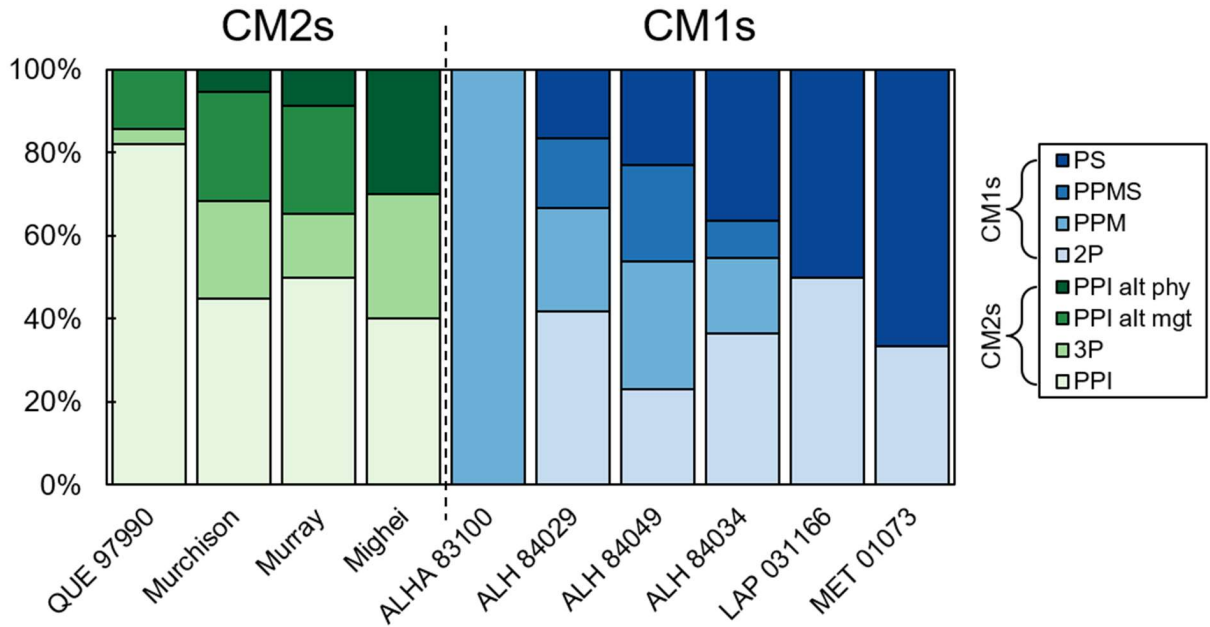

**Figure 10.** Bar plot illustrating the relative proportions of iron sulfide textural groups in CM2 (greens) and CM1 (blues) chondrites. CM2 data are from Singerling and Brearley (2020) with samples listed in order of increasing alteration of the bulk sample moving from left to right. CM1 data are from this study. Note the decrease in PPI grains in the CM2s and the increase in altered sulfides in both groups moving from left to right. These observations are consistent with the following changes in modal abundances: decrease in pyrrhotite and increases in pentlandite, magnetite, and serpentine.

*4.4.4 Comparing alteration conditions among the CM1 iron sulfides*

      According to our observations, $fO_2$, pH, and fluid composition are the primary factors influencing the formation of these different textural groups. Temperature may not have been a significant factor, although it likely plays a role since the difference in alteration temperatures between the CM1 and CM2 chondrites is well established (e.g., Zolensky et al. 1997). The different mineral assemblages produced by alteration of sulfides indicate that they formed under different alteration conditions. How then do we reconcile the variation in alteration conditions for sulfides that are sometimes in close spatial proximity to one another within the same meteorite?

      CM chondrites are commonly brecciated (Metzler et al. 1991), consisting of materials that have been altered to different degrees and have been mixed together by regolith gardening. Therefore, brecciation could have played a role in causing juxtaposition of sulfides with different apparent alteration histories. Alternatively, microchemical environments have been shown to play an important role in the heterogeneity of the samples in regards to alteration conditions. Brearley (2006b) described microchemical environments as localized regions (10s of microns) characterized by small differences in geochemical conditions that cause variations in the style and degree of alteration of primary phases. Palmer and Lauretta (2011) documented different styles of alteration of kamacite in CM chondrites that can be explained by these types of localized variations. CM chondrites are a combination of materials with different compositions, oxidation states, and grain sizes that are heterogeneously mixed on an intimate scale.

Additionally, water, as the major oxidant, was likely heterogeneously distributed from initial accretion as water ice or, if mobile, infiltrated the rocky material heterogeneously. Not surprisingly then, the assemblages resulting from alteration reactions can be variable on a localized scale. Even isotopic signatures, such as carbon from carbonates in CM1 ALH 84049, reflect this localized variability (Tyra et al. 2016). These localized heterogeneities in the conditions of alteration could explain the variable styles of replacement of sulfides that we have observed.

In terms of $fO_2$, the textural groups of altered iron sulfides studied here record alteration under more oxidizing conditions than those under which pyrrhotite is stable but less oxidizing than would permit the formation of phases such as hematite or iron sulfates. The difference in extent of alteration from grain to grain, especially for the formation of magnetite in the PPM grains, could be due to localized differences in $fO_2$ as a result of the heterogeneous availability of water, the primary oxidizing agent.

The proposed alteration reactions also imply differences in pH: more acidic for the 2P grains, evidenced by $H^+$ in the products, and more alkaline for the PPM grains, evidenced by $H^+$ in the reactants. Variations in pH from mildly acidic (pH 6−8; DuFresne and Anders 1962) to moderately alkaline (pH 7–12; Zolensky et al. 1989; Chizmadia and Brearley 2008; Chizmadia et al. 2008) have been documented for the CM chondrites. However, the majority of studies argue for alkaline fluids, based on thermodynamic modeling of mineral assemblages (Zolensky et al. 1989), the hydration reaction of matrix amorphous silicate materials (Chizmadia and Brearley 2008), and the survival of micron-sized Fe,Ni metal in fine-grained rims (Chizmadia et al. 2008). Acidic fluids have been theorized to form by the melting of ices containing HCl hydrates (Zolotov and Mironenko 2007), though this is most pertinent for the earliest stages of alteration. Most likely, localized differences in pH exist due to specific reactions, such as silica-rich chondrule glass reacting with water to form silicic acid (Brearley 2006b). Therefore, although alteration of CM chondrites occurs under alkaline conditions, variations in pH on a localized scale also play an important role in controlling mineral reactions and elemental mobility.

Lastly, the iron sulfides also indicate alteration under different fluid compositions: 1) reaction of pyrrhotite to pentlandite in the 2P and PPM grains requires a Ni-bearing fluid and yields a Fe-bearing fluid; 2) reaction of pentlandite to magnetite in the PPM grains yields a Ni- and S-bearing fluid; and 3) reaction of magnetite to serpentine in the PPM/PS as well as the formation of phyllosilicates in pores of the 2P grains requires a Si- and Mg-bearing fluid. It is not clear exactly what phases would have been involved in controlling the fluid composition in CM1 chondrites, because at their advanced stages of alteration, none of the primary phases are preserved. Forsteritic olivine is the most abundant primary phase in CM chondrites and is the most resistant to alteration (Hanowski and Brearley 2001). It is plausible that the final stages of alteration of forsterite phenocrysts in chondrules controlled the activity of Si and Mg in the fluids. Alternatively, progressive changes in the composition of serpentine, from the more Fe-rich compositions (e.g., cronstedtite) found in CM2 chondrites, to the more Mg-rich compositions in CM1 chondrites, also influenced the fluid composition (Browning et al. 1996; Brearley 2006a).

In summary, the 2P, PPM, PS, and PPMS grains were all initially PPI grains that experienced different alteration conditions, which yielded different alteration products and textures. The alteration assemblages in these different sulfide grains are most likely controlled by alteration of adjacent phases that influenced the local composition and evolution of the fluids significantly. The change in stability from one mineral assemblage to another only requires small changes in geochemical conditions. So, although the changes in conditions may be subtle, the alteration produced can be quite distinct.

## 5. CONCLUSIONS

Similar to observations of iron sulfides in the highly-altered CM2 chondrites, we have demonstrated that primary pyrrhotite in the CM1 chondrites is unstable during aqueous alteration, breaking down into secondary pentlandite, magnetite, and serpentine, whereas primary pentlandite is largely resistant to the alteration. Using a combination of textural and compositional information obtained using several different instruments (FEGSEM/FIB, EPMA, and S/TEM), we determined that the 2P grains are the products of the advanced dissolution of primary pyrrhotite with replacement by secondary pentlandite and precipitation of phyllosilicates in pore space; the PPM grains are the products of oxidation of primary pyrrhotite by a Ni-bearing fluid, resulting in the formation of secondary pentlandite followed by magnetite; the PS grains are likely the products of continued alteration of the PPM grains with a change in fluid composition resulting in the replacement of magnetite by serpentine; and the PPMS grains represent an intermediate step between the PPM and PS grains.

These different textural groups of sulfides all appear to be derived from the primary PPI grains similar to, but not necessarily from, the same population of grains found in the weakly- to moderately-altered CM2 chondrites. They illustrate the complex nature of the aqueous alteration environment that these meteorites were exposed to. There is evidence for changes or fluctuations in pH from the dissolution of pyrrhotite in the 2P grains, in oxygen fugacity from the formation of magnetite, and in fluid compositions from the formation of secondary pentlandite and phyllosilicates, including serpentine. These varying conditions were widespread given the presence of most of these grain types across different meteorites but were controlled on a highly localized scale resulting from distinct microchemical environments.

**Acknowledgements**

We thank Associate Editor Yves Marrocchi, Krysten Villalon, and one anonymous reviewer for excellent feedback and suggestions which improved the manuscript. We also thank Lindsay Keller, Rhian Jones, Elena Dobrica, Jon Lewis, and Mike Spilde for much appreciated advice and recommendations. We acknowledge the Meteorite Working Group for providing the samples used in this study. U.S. Antarctic meteorite samples are recovered by the Antarctic Search for Meteorites (ANSMET) program, which has been funded by NSF and NASA, and characterized and curated by the Department of Mineral Sciences at the Smithsonian Institution and the Astromaterials Curation Office at NASA Johnson Space Center. Electron microscopy and FIB sample preparation were carried out in the Electron Microbeam Analysis Facility in the Department of Earth and Planetary Sciences and the Nanomaterials Characterization Facility, University of New Mexico, which are supported by the State of New Mexico, NSF, and NASA. This work was supported by NASA Cosmochemistry Grant NNX15AD28G and NASA Emerging Worlds Grant 80NSSC18K0731 to A.J. Brearley (PI) Acquisition of the JEOL NEOARM AC-STEM at the University of New Mexico was supported by NSF DMR-1828731.

**Appendices**

Appendix A. Includes mass balance calculations and supplementary figures (Figures A1–A3).
Appendix B. Includes supplementary tables (Tables A1–A5).

# Appendix A to The Fate of Primary Iron Sulfides in the CM1 Carbonaceous Chondrites: Effects of Advanced Aqueous Alteration on Primary Components

S. A. Singerling[1], C. M. Corrigan[2], and A. J. Brearley[1]
[1] Department of Earth and Planetary Sciences, MSC-03 2040 1 University of New Mexico, Albuquerque, NM 87131, USA
[2] Department of Mineral Sciences, National Museum of Natural History, Smithsonian Institution, Washington, DC 20560, USA

## Mass Balance Calculation

For this calculation, we use the following equation:

$$(vol\%_{tochilinit} \times wt\% Ni_{tochilinit}) + (vol\%_{pyrrhotite} \times wt\% Ni_{pyrrhotit})$$
$$= (vol\%_{pentlandite} \times wt\% Ni_{pentlandite})$$

In this equation, tochilinite and pyrrhotite abundances and compositions are for CM2 chondrites, whereas pentlandite abundances and compositions are specifically for secondary (2P grain) pentlandite in CM1 chondrites. The tochilinite modal abundance in CM2 chondrites varies from less than 1 to 7 vol. % (Zolensky et al. 1993; McSween 1987), and the Ni abundance in the tochilinite varies from 3 to 7 wt. % (Palmer and Lauretta 2011). The modal abundance of pyrrhotite in CM2 chondrites varies from 1 to 2 vol. % (Howard et al. 2009; 2011), and the Ni abundance in the pyrrhotite varies from 0.3 to 2.7 wt. % (from ±1 standard deviation of the mean using PPI grain data in Singerling and Brearley, 2018). Lastly, for the Ni abundance in the 2P grain pentlandite, data from the current study is used which varies from 30.7 to 31.8 wt. % (±1 standard deviation of the mean). To account for the range in modal abundances of the phases and Ni contents of those phases, the mass balance calculation was performed using both minimum and maximum values in the ranges.

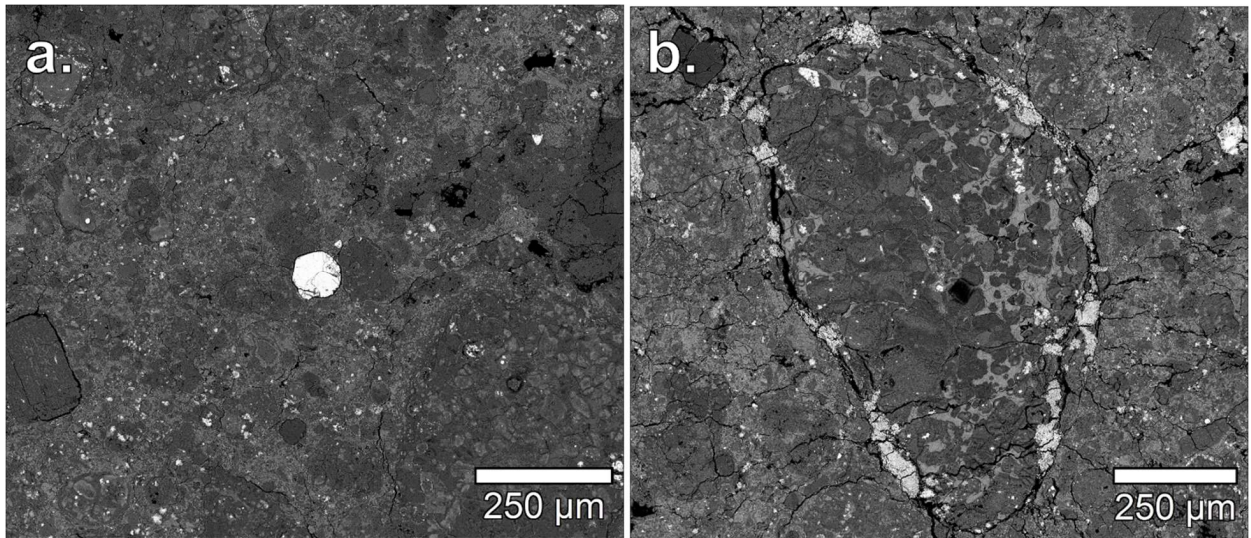

**Figure A1.** BSE images of locations where iron sulfides were identified in CM1 chondrites, including in the matrix and pseudomorphed chondrules. (a) shows an example of the matrix of ALH 84029, with the iron sulfide present in the center of the image, and (b) shows an example of a pseudomorphed chondrule with sulfides on its rim from ALH 84049.

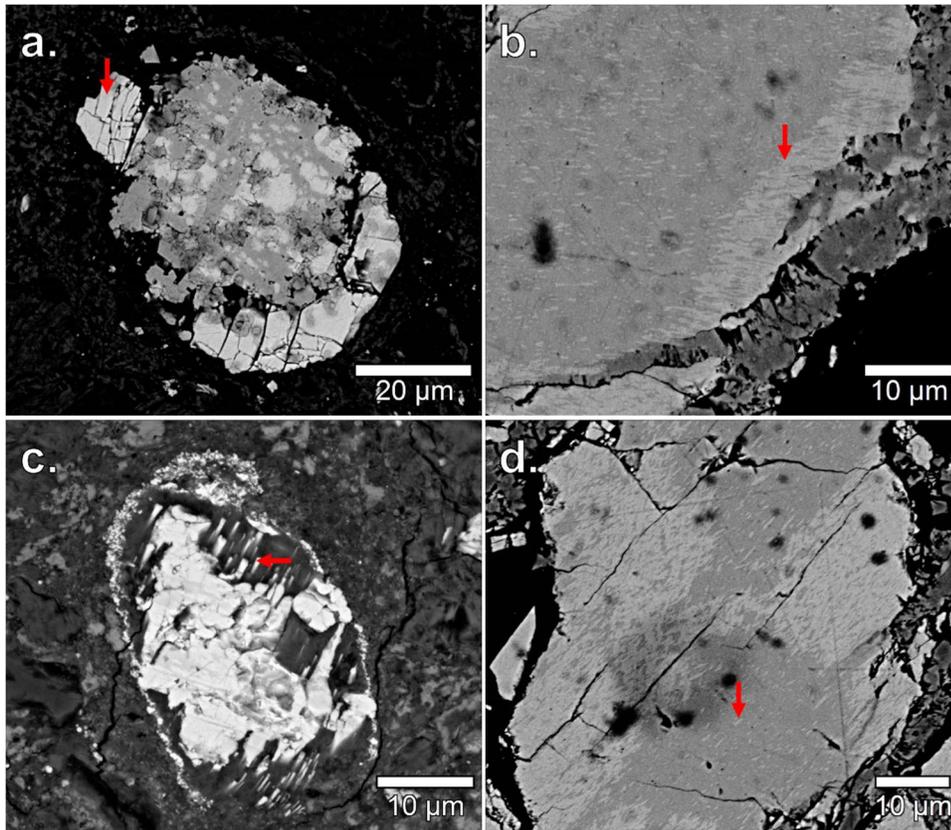

**Figure A2.** BSE images of the common pentlandite exsolution textures in the altered primary sulfides in CM1 chondrites—(a) patches, (b) blades, (c) lamellae, and (d) rods. Red arrows point to examples of each.

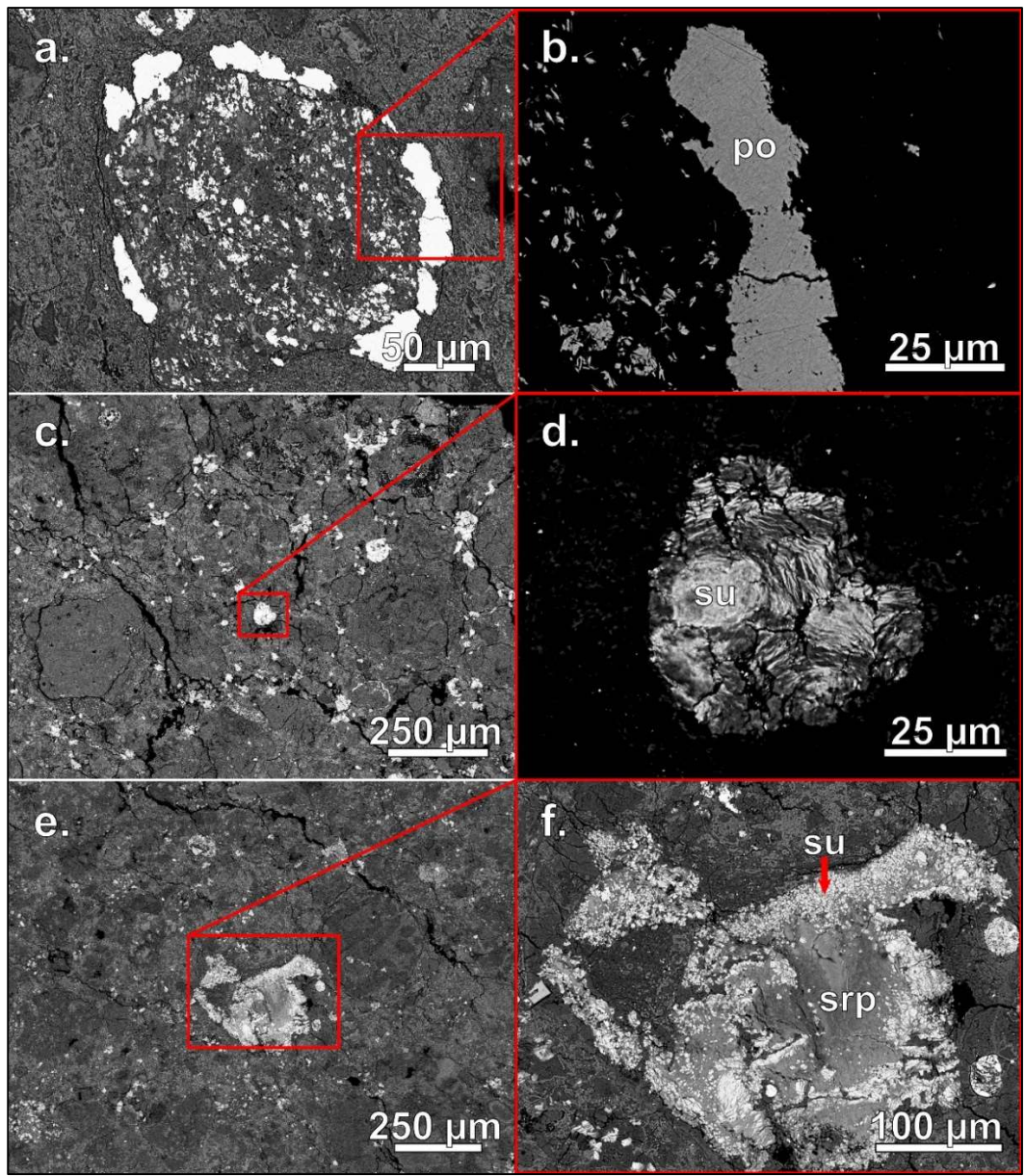

**Figure A3.** BSE images of iron sulfide grains which do not fit into the aforementioned textural groups from CM1 chondrites. They show (a) the textural context of a grain on a pseudomorphed chondrule rim in ALH 84029, (b) a plain pyrrhotite grain with no pentlandite exsolution textures, (c) the textural context of a grain in the a more coarse-grained portion of the matrix in ALH 84029, (d) a coarse-layered sulfide grain with alternating high and low Z phases, (e) the textural context of a grain in the matrix of ALH 84029 surrounded predominantly by serpentine, and (f) a sulfide-rimmed serpentine assemblage with an irregular shape. These grains occur throughout the samples they are observed in (MET 01073 for the plain pyrrhotite and the ALH pairing group for the coarse-layered sulfide and sulfide-rimmed serpentine assemblage). Po = pyrrhotite, su = sulfide, srp = serpentine.

**Table A1.** Elements, crystals, count times, detection limits, and standards for EPMA analyses

| Element | Crystal | Count time (s) | Detection limit (wt. %) | Standard |
|---|---|---|---|---|
| P | PETL | 40 | 0.01–0.03 | GaP |
| S | PETL | 30 | 0.01–0.02 | Pyrite |
| Cr | PETJ | 60–70 | 0.02–0.04 | Chromite |
| Fe | LIF | 40 | 0.05–0.09 | Pyrite |
| Co | LIFH | 40 | 0.02–0.03 | Co metal |
| Ni | LIFH | 40 | 0.02–0.04 | Ni metal |

**Table A2.** Individual phase compositions from EPMA analyses

| Textural group | Grain | Phase | Point | Weight % | | | | | | | Atomic % | | | | | | |
|---|---|---|---|---|---|---|---|---|---|---|---|---|---|---|---|---|---|
| | | | | P | S | Cr | Fe | Co | Ni | Total | P | S | Cr | Fe | Co | Ni | Total |
| PPM | ALH83_S1 | Pn p | 1 | bdl | 32.49 | bdl | 30.68 | 1.05 | 33.59 | 97.8 | bdl | 47.07 | bdl | 25.52 | 0.83 | 26.58 | 100 |
| PPM | ALH83_S1 | Pn p | 2 | bdl | 33.16 | 0.03 | 30.48 | 1.15 | 33.71 | 98.5 | bdl | 47.57 | 0.02 | 25.10 | 0.90 | 26.41 | 100 |
| PPM | ALH83_S1 | Pn b | 3 | bdl | 32.94 | bdl | 33.82 | 0.37 | 31.59 | 98.7 | bdl | 47.18 | bdl | 27.81 | 0.29 | 24.71 | 100 |
| PPM | ALH83_S3 | Pn p | 1 | bdl | 32.80 | bdl | 31.74 | 0.87 | 32.84 | 98.3 | bdl | 47.24 | bdl | 26.24 | 0.69 | 25.83 | 100 |
| PPM | ALH83_S3 | Pn p | 2 | bdl | 32.37 | bdl | 31.87 | 0.85 | 32.73 | 97.8 | bdl | 46.91 | bdl | 26.52 | 0.67 | 25.90 | 100 |
| PPM | ALH83_S4 | Pn b | 1 | bdl | 33.16 | bdl | 35.25 | 0.28 | 29.63 | 98.3 | bdl | 47.55 | bdl | 29.02 | 0.22 | 23.21 | 100 |
| PPM | ALH29_S15 | Pn p | 1 | bdl | 33.29 | bdl | 34.63 | 0.57 | 29.44 | 97.9 | bdl | 47.86 | bdl | 28.58 | 0.44 | 23.11 | 100 |
| PPM | ALH29_S15 | Pn b | 2 | bdl | 32.84 | bdl | 32.63 | 0.51 | 31.59 | 97.6 | bdl | 47.52 | bdl | 27.10 | 0.40 | 24.96 | 100 |
| PPM | ALH49_S1 | Pn b | 1 | bdl | 33.01 | 0.07 | 34.98 | 0.24 | 30.03 | 98.3 | bdl | 47.38 | 0.06 | 28.83 | 0.18 | 23.54 | 100 |
| PPM | ALH49_S6 | Pn p | 1 | bdl | 31.52 | bdl | 30.45 | 1.67 | 33.16 | 96.8 | bdl | 46.33 | bdl | 25.70 | 1.34 | 26.62 | 100 |
| PPM | ALH49_S6 | Pn b | 2 | bdl | 32.36 | bdl | 32.97 | 0.53 | 31.11 | 97.0 | bdl | 47.19 | bdl | 27.60 | 0.42 | 24.77 | 100 |
| PPM | ALH49_S6 | Po | 1 | bdl | 37.46 | bdl | 56.91 | bdl | 2.24 | 96.6 | bdl | 52.48 | bdl | 45.78 | bdl | 1.71 | 100 |
| PPM | LAP_S2 | Po | 1 | bdl | 37.38 | 0.07 | 54.87 | 0.21 | 5.17 | 97.7 | bdl | 52.01 | 0.06 | 43.83 | 0.16 | 3.93 | 100 |
| PPM | LAP_S2 | Po | 2 | bdl | 37.43 | 0.06 | 55.82 | 0.21 | 4.42 | 97.9 | bdl | 51.95 | 0.06 | 44.48 | 0.16 | 3.35 | 100 |
| PPM | LAP_S5 | Pn p | 7 | bdl | 32.72 | 0.04 | 35.90 | 1.01 | 26.03 | 95.7 | bdl | 48.03 | 0.04 | 30.25 | 0.81 | 20.87 | 100 |
| PPM | LAP_S5 | Po | 1 | 0.02 | 37.97 | 0.05 | 58.69 | 0.08 | 1.71 | 98.5 | 0.03 | 52.23 | 0.04 | 46.35 | 0.06 | 1.28 | 100 |
| PPM | LAP_S5 | Po | 2 | bdl | 37.88 | 0.04 | 58.56 | 0.07 | 1.68 | 98.2 | bdl | 52.26 | 0.03 | 46.39 | 0.05 | 1.26 | 100 |
| PPM | LAP_S5 | Po | 3 | bdl | 37.97 | 0.06 | 56.92 | 0.12 | 3.05 | 98.1 | bdl | 52.43 | 0.05 | 45.13 | 0.09 | 2.30 | 100 |
| PPM | LAP_S5 | Po | 4 | bdl | 37.87 | 0.08 | 57.52 | 0.15 | 2.86 | 98.5 | bdl | 52.17 | 0.07 | 45.49 | 0.12 | 2.15 | 100 |
| PPM | LAP_S5 | Po | 5 | bdl | 37.78 | 0.07 | 58.57 | 0.10 | 1.93 | 98.4 | bdl | 52.07 | 0.06 | 46.34 | 0.07 | 1.46 | 100 |
| PPM | LAP_S5 | Po | 6 | bdl | 38.00 | 0.07 | 58.67 | 0.07 | 1.73 | 98.5 | bdl | 52.26 | 0.06 | 46.33 | 0.05 | 1.30 | 100 |
| PS | ALH29_C1S1 | Pn c | 1 | bdl | 32.47 | 0.03 | 30.28 | 0.98 | 33.45 | 97.2 | bdl | 47.28 | 0.02 | 25.31 | 0.78 | 26.60 | 100 |
| PS | ALH29_C1S1 | Pn c | 2 | bdl | 32.53 | bdl | 29.69 | 1.23 | 33.50 | 97.0 | bdl | 47.45 | bdl | 24.86 | 0.98 | 26.69 | 100 |
| PS | ALH29_S2 | Pn c | 1 | bdl | 32.18 | bdl | 29.98 | 0.92 | 33.74 | 96.8 | bdl | 47.10 | bdl | 25.19 | 0.73 | 26.97 | 100 |
| PS | ALH49_S13 | Pn c | 2 | bdl | 32.20 | bdl | 29.91 | 1.29 | 33.35 | 96.8 | bdl | 47.15 | bdl | 25.14 | 1.02 | 26.67 | 100 |
| PS | ALH49_S3 | Pn c | 1 | bdl | 32.31 | bdl | 30.99 | 0.84 | 32.50 | 96.7 | bdl | 47.29 | bdl | 26.04 | 0.67 | 25.98 | 100 |
| PS | ALH49_S3 | Pn c | 2 | bdl | 32.28 | bdl | 31.48 | 0.87 | 32.47 | 97.1 | bdl | 47.08 | bdl | 26.36 | 0.69 | 25.86 | 100 |
| PPMS | ALH29_S20 | Pn b | 1 | bdl | 32.80 | 0.04 | 31.68 | 0.75 | 32.14 | 97.4 | bdl | 47.56 | 0.03 | 26.37 | 0.60 | 25.45 | 100 |
| 2P | ALH29_S13 | Pn | 1 | bdl | 32.62 | bdl | 30.55 | 2.86 | 31.21 | 97.2 | bdl | 47.44 | bdl | 25.51 | 2.26 | 24.79 | 100 |
| 2P | ALH29_S13 | Pn | 2 | bdl | 32.61 | bdl | 29.77 | 2.80 | 31.12 | 96.3 | bdl | 47.79 | bdl | 25.05 | 2.23 | 24.91 | 100 |

Table A2 cont.

| Textural group | Grain | Phase | Point | Weight % | | | | | | | Atomic % | | | | | | |
|---|---|---|---|---|---|---|---|---|---|---|---|---|---|---|---|---|---|
| | | | | P | S | Cr | Fe | Co | Ni | Total | P | S | Cr | Fe | Co | Ni | Total |
| 2P | ALH29_S13 | Pn | 3 | bdl | 32.68 | bdl | 30.10 | 2.80 | 31.35 | 97.0 | bdl | 47.61 | bdl | 25.18 | 2.22 | 24.95 | 100 |
| 2P | ALH29_S13 | Pn | 4 | bdl | 32.75 | bdl | 29.76 | 2.61 | 31.09 | 96.2 | bdl | 48.00 | bdl | 25.03 | 2.08 | 24.88 | 100 |
| 2P | ALH29_S13 | Pn | 6 | bdl | 31.48 | bdl | 29.45 | 1.56 | 32.57 | 95.1 | bdl | 46.96 | bdl | 25.22 | 1.26 | 26.54 | 100 |
| 2P | ALH29_S13 | Pn | 7 | bdl | 32.75 | bdl | 30.09 | 2.85 | 31.44 | 97.2 | bdl | 47.63 | bdl | 25.12 | 2.25 | 24.97 | 100 |
| 2P | ALH29_S13 | Pn | 8 | bdl | 31.80 | bdl | 30.03 | 2.65 | 30.70 | 95.2 | bdl | 47.27 | bdl | 25.63 | 2.14 | 24.93 | 100 |
| 2P | ALH29_S13 | Pn | 9 | bdl | 32.71 | bdl | 30.10 | 2.71 | 31.15 | 96.7 | bdl | 47.75 | bdl | 25.23 | 2.16 | 24.84 | 100 |
| 2P | ALH29_S6 | Pn | 1 | bdl | 32.55 | bdl | 30.46 | 2.53 | 31.11 | 96.7 | bdl | 47.56 | bdl | 25.56 | 2.01 | 24.83 | 100 |
| 2P | ALH29_S6 | Pn | 3 | bdl | 32.78 | bdl | 29.99 | 2.62 | 31.60 | 97.0 | bdl | 47.73 | bdl | 25.06 | 2.08 | 25.12 | 100 |
| 2P | ALH29_S9 | Pn | 1 | bdl | 32.55 | bdl | 30.33 | 2.49 | 30.84 | 96.2 | bdl | 47.75 | bdl | 25.54 | 1.98 | 24.70 | 100 |
| 2P | ALH29_S9 | Pn | 5 | bdl | 32.56 | 0.04 | 30.65 | 2.45 | 30.70 | 96.4 | bdl | 47.69 | 0.04 | 25.77 | 1.95 | 24.56 | 100 |
| 2P | ALH34_S1 | Pn | 1 | bdl | 32.29 | 0.05 | 30.23 | 2.55 | 30.69 | 95.8 | bdl | 47.60 | 0.05 | 25.59 | 2.05 | 24.71 | 100 |
| 2P | ALH34_S1 | Pn | 2 | bdl | 32.38 | bdl | 30.35 | 2.51 | 30.70 | 96.0 | bdl | 47.65 | bdl | 25.64 | 2.01 | 24.67 | 100 |
| 2P | ALH34_S1 | Pn | 3 | bdl | 32.52 | 0.04 | 30.14 | 2.60 | 30.72 | 96.0 | bdl | 47.80 | 0.04 | 25.43 | 2.08 | 24.65 | 100 |
| 2P | ALH34_S1 | Pn | 5 | bdl | 32.67 | bdl | 30.88 | 2.47 | 30.92 | 97.0 | bdl | 47.58 | bdl | 25.83 | 1.96 | 24.60 | 100 |
| 2P | ALH34_S13 | Pn | 3 | bdl | 32.28 | bdl | 28.76 | 2.57 | 31.45 | 95.1 | bdl | 47.92 | bdl | 24.51 | 2.07 | 25.49 | 100 |
| 2P | ALH34_S13 | Pn | 6 | bdl | 32.21 | bdl | 29.03 | 2.44 | 31.54 | 95.2 | bdl | 47.75 | bdl | 24.71 | 1.97 | 25.54 | 100 |
| 2P | ALH34_S4 | Pn | 2 | bdl | 32.81 | 0.06 | 30.93 | 2.33 | 30.49 | 96.6 | bdl | 47.87 | 0.06 | 25.91 | 1.85 | 24.30 | 100 |
| 2P | ALH34_S4 | Pn | 4 | bdl | 32.91 | 0.05 | 30.92 | 2.42 | 30.70 | 97.0 | bdl | 47.85 | 0.04 | 25.81 | 1.91 | 24.37 | 100 |
| 2P | ALH49_S5 | Pn | 1 | bdl | 32.81 | bdl | 30.95 | 2.69 | 30.81 | 97.3 | bdl | 47.63 | bdl | 25.80 | 2.13 | 24.43 | 100 |
| 2P | ALH49_S5 | Pn | 3 | bdl | 32.59 | bdl | 30.76 | 2.70 | 30.77 | 96.8 | bdl | 47.55 | bdl | 25.77 | 2.14 | 24.52 | 100 |
| 2P | ALH49_S5 | Pn | 4 | bdl | 32.90 | bdl | 31.08 | 2.87 | 31.08 | 98.0 | bdl | 47.48 | bdl | 25.75 | 2.25 | 24.50 | 100 |
| 2P | ALH49_S5 | Pn | 5 | bdl | 32.38 | bdl | 30.97 | 2.70 | 30.45 | 96.5 | bdl | 47.42 | bdl | 26.04 | 2.15 | 24.35 | 100 |
| 2P | ALH49_S5 | Pn | 6 | bdl | 32.22 | bdl | 30.84 | 2.65 | 30.29 | 96.0 | bdl | 47.43 | bdl | 26.07 | 2.12 | 24.36 | 100 |
| 2P | ALH49_S5 | Pn | 7 | bdl | 32.21 | bdl | 30.54 | 2.68 | 30.54 | 96.0 | bdl | 47.44 | bdl | 25.83 | 2.15 | 24.57 | 100 |
| 2P | ALH49_S5 | Pn | 8 | bdl | 31.94 | bdl | 30.56 | 2.83 | 30.33 | 95.7 | bdl | 47.24 | bdl | 25.95 | 2.28 | 24.50 | 100 |
| 2P | ALH49_S5 | Pn | 9 | bdl | 32.88 | bdl | 30.65 | 3.00 | 31.25 | 97.8 | bdl | 47.53 | bdl | 25.44 | 2.36 | 24.67 | 100 |
| 2P | ALH49_S5 | Pn | 10 | bdl | 32.52 | 0.04 | 30.03 | 3.15 | 30.70 | 96.4 | bdl | 47.64 | 0.03 | 25.26 | 2.51 | 24.56 | 100 |
| 2P | ALH49_S5 | Pn | 11 | bdl | 32.48 | bdl | 30.23 | 2.97 | 30.77 | 96.5 | bdl | 47.58 | bdl | 25.43 | 2.36 | 24.62 | 100 |
| 2P | ALH49_S5 | Pn | 12 | bdl | 32.58 | bdl | 30.88 | 2.83 | 31.14 | 97.5 | bdl | 47.30 | bdl | 25.74 | 2.23 | 24.70 | 100 |
| 2P | ALH49_S7 | Pn | 1 | bdl | 33.08 | bdl | 29.99 | 2.94 | 32.03 | 98.0 | bdl | 47.67 | bdl | 24.81 | 2.31 | 25.21 | 100 |

**Table A2 cont.**

| Textural group | Grain | Phase | Point | Weight % | | | | | | | Atomic % | | | | | | |
|---|---|---|---|---|---|---|---|---|---|---|---|---|---|---|---|---|---|
| | | | | P | S | Cr | Fe | Co | Ni | Total | P | S | Cr | Fe | Co | Ni | Total |
| 2P | ALH49_S7 | Pn | 2 | bdl | 32.42 | bdl | 29.78 | 2.79 | 31.49 | 96.5 | bdl | 47.51 | bdl | 25.06 | 2.22 | 25.21 | 100 |
| 2P | ALH49_S7 | Pn | 3 | bdl | 33.04 | 0.04 | 30.20 | 2.91 | 31.70 | 97.9 | bdl | 47.67 | 0.04 | 25.02 | 2.29 | 24.98 | 100 |
| 2P | ALH49_S7 | Pn | 4 | bdl | 32.90 | bdl | 30.25 | 2.93 | 31.66 | 97.8 | bdl | 47.57 | bdl | 25.11 | 2.31 | 25.00 | 100 |
| 2P | ALH49_S7 | Pn | 5 | bdl | 32.96 | bdl | 30.33 | 2.98 | 32.08 | 98.4 | bdl | 47.41 | bdl | 25.04 | 2.33 | 25.20 | 100 |
| 2P | ALH49_S7 | Pn | 6 | bdl | 32.73 | bdl | 30.18 | 2.80 | 31.40 | 97.1 | bdl | 47.61 | bdl | 25.21 | 2.22 | 24.95 | 100 |
| 2P | ALH49_S7 | Pn | 7 | bdl | 32.73 | bdl | 30.02 | 2.94 | 31.55 | 97.3 | bdl | 47.57 | bdl | 25.05 | 2.33 | 25.04 | 100 |
| 2P | ALH49_S7 | Pn | 8 | bdl | 32.47 | bdl | 30.05 | 2.58 | 31.21 | 96.3 | bdl | 47.62 | bdl | 25.31 | 2.06 | 25.00 | 100 |
| 2P | ALH49_S7 | Pn | 9 | bdl | 32.69 | bdl | 29.89 | 2.76 | 31.90 | 97.3 | bdl | 47.53 | bdl | 24.95 | 2.19 | 25.32 | 100 |
| 2P | ALH49_S7 | Pn | 10 | bdl | 32.95 | 0.03 | 29.98 | 2.95 | 31.64 | 97.6 | bdl | 47.71 | 0.03 | 24.92 | 2.32 | 25.02 | 100 |
| 2P | ALH49_S7 | Pn | 11 | bdl | 32.20 | 0.04 | 29.77 | 2.80 | 30.86 | 95.7 | bdl | 47.57 | 0.03 | 25.25 | 2.25 | 24.90 | 100 |
| 2P | ALH49_S7 | Pn | 12 | bdl | 32.75 | 0.05 | 29.77 | 2.80 | 31.68 | 97.1 | bdl | 47.67 | 0.05 | 24.87 | 2.21 | 25.19 | 100 |
| 2P | LAP_S9 | Pn | 1 | bdl | 32.62 | 0.04 | 30.76 | 2.11 | 32.04 | 97.6 | bdl | 47.31 | 0.04 | 25.61 | 1.67 | 25.38 | 100 |
| 2P | LAP_S9 | Pn | 2 | bdl | 32.33 | bdl | 30.45 | 2.07 | 31.35 | 96.2 | bdl | 47.49 | bdl | 25.68 | 1.65 | 25.15 | 100 |
| 2P | LAP_S9 | Pn | 3 | bdl | 32.21 | 0.05 | 30.24 | 2.07 | 31.85 | 96.4 | bdl | 47.28 | 0.05 | 25.48 | 1.65 | 25.54 | 100 |
| 2P | LAP_S9 | Pn | 4 | bdl | 32.47 | bdl | 30.45 | 2.21 | 31.98 | 97.1 | bdl | 47.31 | bdl | 25.47 | 1.75 | 25.45 | 100 |
| 2P | LAP_S9 | Pn | 5 | bdl | 32.65 | bdl | 30.66 | 2.25 | 32.32 | 97.9 | bdl | 47.23 | bdl | 25.46 | 1.77 | 25.53 | 100 |
| 2P | LAP_S9 | Pn | 6 | bdl | 32.88 | 0.05 | 30.82 | 2.21 | 32.23 | 98.2 | bdl | 47.37 | 0.05 | 25.49 | 1.74 | 25.36 | 100 |
| 2P | LAP_S9 | Pn | 7 | bdl | 32.28 | bdl | 30.99 | 2.07 | 31.70 | 97.1 | bdl | 47.10 | bdl | 25.97 | 1.65 | 25.26 | 100 |

PPM = pyrrhotite-pentlandite-magnetite, PS = pentlandite-serpentine, PPMS = pyrrhotite-pentlandite-magnetite-serpentine, 2P = porous pentlandite, ALH83 = ALHA 83100, ALH29 = ALH 84029, ALH49 = ALH 84049, LAP = LAP 031166, ALH34 = ALH 84034, S# = matrix sulfide #, C#S# = sulfide # in chondrule #, pn = pentlandite, p = patch texture, b = blade texture, c = coarse-grained, po = pyrrhotite, bdl = below detection limit.

**Table A3.** Individual spot and bulk 2P grain compositions (in wt. %) from EPMA analyses

| Grain | Point[1] | S | Cr | Fe | Co | Ni | Total |
|---|---|---|---|---|---|---|---|
| ALH29_S13 | 1 | 32.62 | bdl | 30.55 | 2.86 | 31.21 | 97.2 |
| ALH29_S13 | 2 | 32.61 | bdl | 29.77 | 2.80 | 31.12 | 96.3 |
| ALH29_S13 | 3 | 32.68 | bdl | 30.10 | 2.80 | 31.35 | 97.0 |
| ALH29_S13 | 4 | 32.75 | bdl | 29.76 | 2.61 | 31.09 | 96.2 |
| ALH29_S13 | 6 | 31.48 | bdl | 29.45 | 1.56 | 32.57 | 95.1 |
| ALH29_S13 | 7 | 32.75 | bdl | 30.09 | 2.85 | 31.44 | 97.2 |
| ALH29_S13 | 8 | 31.8 | bdl | 30.03 | 2.65 | 30.70 | 95.2 |
| ALH29_S13 | 9 | 32.71 | bdl | 30.10 | 2.71 | 31.15 | 96.7 |
| ALH29_S13 | Average | 32.43 | bdl | 29.98 | 2.60 | 31.33 | 96.3 |
| ALH29_S13 | Standard deviation | 0.49 | bdl | 0.33 | 0.43 | 0.55 | |
| ALH29_S6 | 1 | 32.55 | bdl | 30.46 | 2.53 | 31.11 | 96.7 |
| ALH29_S6 | 3 | 32.78 | bdl | 29.99 | 2.62 | 31.60 | 97.0 |
| ALH29_S6 | Average | 32.66 | bdl | 30.22 | 2.57 | 31.36 | 96.8 |
| ALH29_S6 | Standard deviation | 0.17 | bdl | 0.34 | 0.07 | 0.34 | |
| ALH29_S9 | 1 | 32.55 | bdl | 30.33 | 2.49 | 30.84 | 96.2 |
| ALH29_S9 | 5 | 32.56 | 0.04 | 30.65 | 2.45 | 30.70 | 96.4 |
| ALH29_S9 | Average | 32.56 | bdl | 30.49 | 2.47 | 30.77 | 96.3 |
| ALH29_S9 | Standard deviation | 0.01 | bdl | 0.23 | 0.03 | 0.09 | |
| ALH34_S1 | 1 | 32.29 | 0.05 | 30.23 | 2.55 | 30.69 | 95.8 |
| ALH34_S1 | 2 | 32.38 | bdl | 30.35 | 2.51 | 30.70 | 96.0 |
| ALH34_S1 | 3 | 32.52 | 0.04 | 30.14 | 2.60 | 30.72 | 96.0 |
| ALH34_S1 | 5 | 32.67 | bdl | 30.88 | 2.47 | 30.92 | 97.0 |
| ALH34_S1 | Average | 32.46 | 0.05 | 30.40 | 2.53 | 30.76 | 96.2 |
| ALH34_S1 | Standard deviation | 0.17 | 0.01 | 0.33 | 0.05 | 0.11 | |
| ALH34_S13 | 3 | 32.28 | bdl | 28.76 | 2.57 | 31.45 | 95.1 |
| ALH34_S13 | 6 | 32.21 | bdl | 29.03 | 2.44 | 31.54 | 95.2 |
| ALH34_S13 | Average | 32.25 | bdl | 28.89 | 2.50 | 31.49 | 95.1 |
| ALH34_S13 | Standard deviation | 0.05 | bdl | 0.19 | 0.09 | 0.07 | |
| ALH34_S4 | 2 | 32.81 | 0.06 | 30.93 | 2.33 | 30.49 | 96.6 |
| ALH34_S4 | 4 | 32.91 | 0.05 | 30.92 | 2.42 | 30.70 | 97.0 |
| ALH34_S4 | Average | 32.86 | 0.05 | 30.92 | 2.37 | 30.59 | 96.8 |
| ALH34_S4 | Standard deviation | 0.07 | 0.01 | 0.01 | 0.06 | 0.14 | |

Table A3 cont.

| Grain | Point[1] | S | Cr | Fe | Co | Ni | Total |
|---|---|---|---|---|---|---|---|
| ALH49_S5 | 1 | 32.81 | bdl | 30.95 | 2.69 | 30.81 | 97.3 |
| ALH49_S5 | 3 | 32.59 | bdl | 30.76 | 2.70 | 30.77 | 96.8 |
| ALH49_S5 | 4 | 32.9 | bdl | 31.08 | 2.87 | 31.08 | 98.0 |
| ALH49_S5 | 5 | 32.38 | bdl | 30.97 | 2.70 | 30.45 | 96.5 |
| ALH49_S5 | 6 | 32.22 | bdl | 30.84 | 2.65 | 30.29 | 96.0 |
| ALH49_S5 | 7 | 32.21 | bdl | 30.54 | 2.68 | 30.54 | 96.0 |
| ALH49_S5 | 8 | 31.94 | bdl | 30.56 | 2.83 | 30.33 | 95.7 |
| ALH49_S5 | 9 | 32.88 | bdl | 30.65 | 3.00 | 31.25 | 97.8 |
| ALH49_S5 | 10 | 32.52 | 0.04 | 30.03 | 3.15 | 30.70 | 96.4 |
| ALH49_S5 | 11 | 32.48 | bdl | 30.23 | 2.97 | 30.77 | 96.5 |
| ALH49_S5 | 12 | 32.58 | bdl | 30.88 | 2.83 | 31.14 | 97.5 |
| ALH49_S5 | Average | 32.5 | bdl | 30.68 | 2.82 | 30.74 | 96.7 |
| ALH49_S5 | Standard deviation | 0.3 | bdl | 0.32 | 0.16 | 0.32 | |
| ALH49_S7 | 1 | 33.08 | bdl | 29.99 | 2.94 | 32.03 | 98.0 |
| ALH49_S7 | 2 | 32.42 | bdl | 29.78 | 2.79 | 31.49 | 96.5 |
| ALH49_S7 | 3 | 33.04 | 0.04 | 30.20 | 2.91 | 31.70 | 97.9 |
| ALH49_S7 | 4 | 32.9 | bdl | 30.25 | 2.93 | 31.66 | 97.8 |
| ALH49_S7 | 5 | 32.96 | bdl | 30.33 | 2.98 | 32.08 | 98.4 |
| ALH49_S7 | 6 | 32.73 | bdl | 30.18 | 2.80 | 31.40 | 97.1 |
| ALH49_S7 | 7 | 32.73 | bdl | 30.02 | 2.94 | 31.55 | 97.3 |
| ALH49_S7 | 8 | 32.47 | bdl | 30.05 | 2.58 | 31.21 | 96.3 |
| ALH49_S7 | 9 | 32.69 | bdl | 29.89 | 2.76 | 31.90 | 97.3 |
| ALH49_S7 | 10 | 32.95 | 0.03 | 29.98 | 2.95 | 31.64 | 97.6 |
| ALH49_S7 | 11 | 32.2 | 0.04 | 29.77 | 2.80 | 30.86 | 95.7 |
| ALH49_S7 | 12 | 32.75 | 0.05 | 29.77 | 2.80 | 31.68 | 97.1 |
| ALH49_S7 | Average | 32.74 | 0.04 | 30.02 | 2.85 | 31.60 | 97.3 |
| ALH49_S7 | Standard deviation | 0.27 | 0.01 | 0.19 | 0.11 | 0.34 | |
| LAP_S9 | 1 | 32.62 | 0.04 | 30.76 | 2.11 | 32.04 | 97.6 |
| LAP_S9 | 2 | 32.33 | bdl | 30.45 | 2.07 | 31.35 | 96.2 |
| LAP_S9 | 3 | 32.21 | 0.05 | 30.24 | 2.07 | 31.85 | 96.4 |
| LAP_S9 | 4 | 32.47 | bdl | 30.45 | 2.21 | 31.98 | 97.1 |
| LAP_S9 | 5 | 32.65 | bdl | 30.66 | 2.25 | 32.32 | 97.9 |
| LAP_S9 | 6 | 32.88 | 0.05 | 30.82 | 2.21 | 32.23 | 98.2 |
| LAP_S9 | 7 | 32.28 | bdl | 30.99 | 2.07 | 31.70 | 97.1 |
| LAP_S9 | Average | 32.49 | 0.05 | 30.62 | 2.14 | 31.92 | 97.2 |
| LAP_S9 | Standard deviation | 0.24 | N/A | 0.26 | 0.08 | 0.33 | |

[1]Point numbers does not necessarily match the point numbers in Table A2, as the data here are a subset.

ALH83 = ALHA 83100, ALH29 = ALH 84029, ALH49 = ALH 84049, LAP = LAP 031166, ALH34 = ALH 84034, S# = matrix sulfide #, bdl = below detection limit, N/A = not

**Table A4.** CM1 ALH 84049 PS grain S13 serpentine compositional data from TEM EDS analyses

| Texture | Point | Weight % | | | | | Atomic % | | | | | Mg#[1] |
|---|---|---|---|---|---|---|---|---|---|---|---|---|
| | | O | Mg | Si | Fe | Total | O | Mg | Si | Fe | Total | |
| Smooth | 1 | 57.4 | 16.2 | 15.3 | 11.1 | 100 | 71.8 | 13.3 | 10.9 | 4.0 | 100 | 0.77 |
| Smooth | 2 | 58.5 | 16.1 | 16.0 | 9.50 | 100 | 72.3 | 13.1 | 11.3 | 3.4 | 100 | 0.80 |
| Smooth | 3 | 58.2 | 14.9 | 16.2 | 10.7 | 100 | 72.5 | 12.2 | 11.5 | 3.8 | 100 | 0.76 |
| Smooth | 4 | 56.4 | 16.1 | 16.5 | 11.0 | 100 | 70.9 | 13.3 | 11.8 | 4.0 | 100 | 0.77 |
| Smooth | Average | 57.6 | 15.8 | 16.0 | 10.6 | | 71.9 | 13.0 | 11.4 | 3.8 | | 0.77 |
| Smooth | Standard deviation | 0.9 | 0.6 | 0.5 | 0.7 | | 0.7 | 0.5 | 0.4 | 0.3 | | 0.01 |
| Fibrous | 1 | 60.1 | 16.8 | 16.5 | 6.6 | 100 | 72.9 | 13.4 | 11.4 | 2.3 | 100 | 0.85 |
| Fibrous | 2 | 60.3 | 15.9 | 16.4 | 7.4 | 100 | 73.3 | 12.7 | 11.4 | 2.6 | 100 | 0.83 |
| Fibrous | 3 | 62.1 | 12.9 | 17.6 | 7.4 | 100 | 75.1 | 10.3 | 12.1 | 2.6 | 100 | 0.80 |
| Fibrous | 4 | 60.2 | 15.5 | 16.7 | 7.5 | 100 | 73.3 | 12.4 | 11.6 | 2.6 | 100 | 0.83 |
| Fibrous | Average | 60.7 | 15.3 | 16.8 | 7.2 | | 73.7 | 12.2 | 11.6 | 2.5 | | 0.83 |
| Fibrous | Standard deviation | 1.0 | 1.7 | 0.5 | 0.4 | | 1.0 | 1.4 | 0.3 | 0.1 | | 0.04 |

[1]Mg# = Mg/(Mg+Fe) in atomic %

**Table A5.** CM1 and CM2[1] pyrrhotite compositional data from EPMA analyses used to calculate the Fe/S and (Fe+Co+Ni+Cr)/S ratios

| Meteorite group | Textural group | Grain | Point | Weight % | | | | | | | Atomic % | | | | | | | Atomic ratio | |
|---|---|---|---|---|---|---|---|---|---|---|---|---|---|---|---|---|---|---|---|
| | | | | P | S | Cr | Fe | Co | Ni | Total | P | S | Cr | Fe | Co | Ni | Total | Fe/S | (Fe+Co+Ni+Cr)/S |
| CM1 | PPM | ALH49_S6 | 1 | bdl | 37.46 | bdl | 56.91 | bdl | 2.24 | 96.6 | bdl | 52.48 | bdl | 45.78 | bdl | 1.71 | 100 | 0.87 | 0.90 |
| CM1 | PPM | LAP_S2 | 1 | bdl | 37.38 | 0.07 | 54.87 | 0.21 | 5.17 | 97.7 | bdl | 52.01 | 0.06 | 43.83 | 0.16 | 3.93 | 100 | 0.84 | 0.92 |
| CM1 | PPM | LAP_S2 | 2 | bdl | 37.43 | 0.06 | 55.82 | 0.21 | 4.42 | 97.9 | bdl | 51.95 | 0.06 | 44.48 | 0.16 | 3.35 | 100 | 0.86 | 0.92 |
| CM1 | PPM | LAP_S5 | 1 | 0.02 | 37.97 | 0.05 | 58.69 | 0.08 | 1.71 | 98.5 | 0.031 | 52.23 | 0.04 | 46.35 | 0.06 | 1.28 | 100 | 0.89 | 0.91 |
| CM1 | PPM | LAP_S5 | 2 | bdl | 37.88 | 0.04 | 58.56 | 0.07 | 1.68 | 98.2 | bdl | 52.26 | 0.03 | 46.39 | 0.05 | 1.26 | 100 | 0.89 | 0.91 |
| CM1 | PPM | LAP_S5 | 3 | bdl | 37.97 | 0.06 | 56.92 | 0.12 | 3.05 | 98.1 | bdl | 52.43 | 0.05 | 45.13 | 0.09 | 2.30 | 100 | 0.86 | 0.91 |
| CM1 | PPM | LAP_S5 | 4 | bdl | 37.87 | 0.08 | 57.52 | 0.15 | 2.86 | 98.5 | bdl | 52.17 | 0.07 | 45.49 | 0.12 | 2.15 | 100 | 0.87 | 0.92 |
| CM1 | PPM | LAP_S5 | 5 | bdl | 37.78 | 0.07 | 58.57 | 0.1 | 1.93 | 98.4 | bdl | 52.07 | 0.06 | 46.34 | 0.07 | 1.46 | 100 | 0.89 | 0.92 |
| CM1 | PPM | LAP_S5 | 6 | bdl | 38 | 0.07 | 58.67 | 0.07 | 1.73 | 98.5 | bdl | 52.26 | 0.06 | 46.33 | 0.05 | 1.30 | 100 | 0.89 | 0.91 |
| CM2 | PPI alt mgt | QUE97_C18S1a | 2 | na | 36.83 | 0.04 | 62.66 | 0.06 | 0.56 | 100.1 | na | 50.34 | 0.04 | 49.16 | 0.04 | 0.42 | 100 | 0.98 | 0.99 |
| CM2 | PPI alt mgt | QUE97_C18S1a | 3 | na | 36.13 | 0.04 | 59.50 | 0.31 | 4.54 | 100.5 | na | 49.52 | 0.04 | 46.82 | 0.23 | 3.40 | 100 | 0.95 | 1.02 |
| CM2 | PPI alt mgt | QUE97_C18S1a | 4 | na | 36.68 | 0.04 | 63.04 | 0.09 | 0.55 | 100.4 | na | 50.08 | 0.03 | 49.41 | 0.07 | 0.41 | 100 | 0.99 | 1.00 |
| CM2 | PPI alt mgt | QUE97_C18S1a | 7 | na | 35.72 | 0.03 | 56.95 | 0.31 | 6.94 | 99.9 | na | 49.35 | 0.02 | 45.16 | 0.23 | 5.23 | 100 | 0.92 | 1.03 |
| CM2 | PPI alt mgt | QUE97_C18S1a | 8 | na | 37.00 | 0.05 | 62.63 | 0.06 | 0.84 | 100.6 | na | 50.36 | 0.04 | 48.93 | 0.05 | 0.62 | 100 | 0.97 | 0.99 |
| CM2 | PPI alt mgt | QUE97_C18S1b | 1 | na | 36.96 | 0.04 | 63.12 | 0.06 | 0.39 | 100.6 | na | 50.31 | 0.03 | 49.32 | 0.04 | 0.29 | 100 | 0.98 | 0.99 |
| CM2 | PPI alt mgt | QUE97_C18S1b | 2 | na | 36.93 | 0.05 | 63.15 | 0.04 | 0.39 | 100.6 | na | 50.28 | 0.04 | 49.36 | 0.03 | 0.29 | 100 | 0.98 | 0.99 |
| CM2 | PPI alt mgt | QUE97_C18S1b | 3 | na | 36.86 | 0.05 | 62.83 | 0.05 | 0.35 | 100.1 | na | 50.37 | 0.04 | 49.29 | 0.04 | 0.26 | 100 | 0.98 | 0.99 |
| CM2 | PPI alt mgt | QUE97_C18S1b | 4 | na | 36.31 | 0.04 | 62.19 | 0.08 | 0.53 | 99.2 | na | 50.17 | 0.03 | 49.33 | 0.06 | 0.40 | 100 | 0.98 | 0.99 |
| CM2 | PPI alt mgt | QUE97_C18S1b | 5 | na | 36.91 | 0.05 | 63.28 | 0.10 | 0.32 | 100.7 | na | 50.23 | 0.04 | 49.42 | 0.07 | 0.24 | 100 | 0.98 | 0.99 |
| CM2 | PPI alt mgt | QUE97_C18S1b | 6 | na | 37.07 | 0.04 | 63.12 | 0.06 | 0.37 | 100.7 | na | 50.39 | 0.03 | 49.26 | 0.04 | 0.28 | 100 | 0.98 | 0.98 |
| CM2 | PPI alt mgt | QUE97_C18S1b | 7 | na | 37.05 | 0.05 | 63.04 | 0.05 | 0.45 | 100.7 | na | 50.39 | 0.04 | 49.21 | 0.03 | 0.33 | 100 | 0.98 | 0.98 |
| CM2 | PPI alt mgt | QUE97_S5 | 1 | na | 36.69 | 0.01 | 62.38 | 0.18 | 1.21 | 100.5 | na | 50.08 | 0.01 | 48.87 | 0.14 | 0.90 | 100 | 0.98 | 1.00 |
| CM2 | PPI alt mgt | QUE97_S5 | 2 | na | 36.58 | 0.01 | 61.64 | 0.19 | 1.64 | 100.1 | na | 50.13 | 0.01 | 48.49 | 0.14 | 1.23 | 100 | 0.97 | 0.99 |
| CM2 | PPI alt mgt | QUE97_S5 | 3 | na | 36.94 | 0.02 | 61.92 | 0.20 | 1.61 | 100.7 | na | 50.27 | 0.01 | 48.37 | 0.15 | 1.20 | 100 | 0.96 | 0.99 |
| CM2 | PPI alt mgt | QUE97_S5 | 4 | na | 36.79 | 0.02 | 61.37 | 0.20 | 1.80 | 100.2 | na | 50.31 | 0.01 | 48.18 | 0.15 | 1.35 | 100 | 0.96 | 0.99 |
| CM2 | PPI alt mgt | QUE97_S5 | 5 | na | 36.86 | 0.01 | 61.55 | 0.22 | 1.75 | 100.4 | na | 50.31 | 0.01 | 48.22 | 0.16 | 1.30 | 100 | 0.96 | 0.99 |
| CM2 | PPI alt mgt | QUE97_S5 | 6 | na | 35.96 | 0.01 | 61.41 | 0.21 | 1.62 | 99.2 | na | 49.80 | 0.00 | 48.81 | 0.16 | 1.23 | 100 | 0.98 | 1.01 |
| CM2 | PPI alt mgt | QUE97_S5 | 7 | na | 36.39 | 0.02 | 61.25 | 0.21 | 1.69 | 99.6 | na | 50.13 | 0.02 | 48.43 | 0.15 | 1.27 | 100 | 0.97 | 0.99 |
| CM2 | PPI alt mgt | QUE97_S5 | 8 | na | 36.48 | 0.02 | 61.91 | 0.20 | 1.64 | 100.2 | na | 49.95 | 0.01 | 48.66 | 0.15 | 1.22 | 100 | 0.97 | 1.00 |
| CM2 | PPI alt mgt | QUE97_S5 | 9 | na | 36.41 | 0.02 | 61.76 | 0.22 | 1.68 | 100.1 | na | 49.94 | 0.02 | 48.62 | 0.16 | 1.26 | 100 | 0.97 | 1.00 |
| CM2 | PPI | QUE97_C6S1 | 1 | na | 36.46 | 0.02 | 62.99 | 0.14 | 0.91 | 100.5 | na | 49.80 | 0.02 | 49.39 | 0.11 | 0.68 | 100 | 0.99 | 1.01 |
| CM2 | PPI | QUE97_C6S1 | 2 | na | 35.74 | 0.04 | 59.07 | 0.19 | 4.48 | 99.5 | na | 49.48 | 0.03 | 46.96 | 0.14 | 3.38 | 100 | 0.95 | 1.02 |
| CM2 | PPI | QUE97_C6S1 | 4 | na | 36.28 | 0.03 | 62.66 | 0.13 | 1.06 | 100.2 | na | 49.76 | 0.02 | 49.33 | 0.10 | 0.79 | 100 | 0.99 | 1.01 |
| CM2 | PPI | QUE97_C6S1 | 5 | na | 36.59 | 0.03 | 62.90 | 0.10 | 0.74 | 100.4 | na | 50.00 | 0.03 | 49.34 | 0.08 | 0.55 | 100 | 0.99 | 1.00 |

Table A5 cont.

| Meteorite group | Textural group | Grain | Point | Weight % | | | | | | | Atomic % | | | | | | | Atomic ratio | |
|---|---|---|---|---|---|---|---|---|---|---|---|---|---|---|---|---|---|---|---|
| | | | | P | S | Cr | Fe | Co | Ni | Total | P | S | Cr | Fe | Co | Ni | Total | Fe/S | (Fe+Co+Ni+Cr)/S |
| CM2 | PPI | QUE97_C6S1 | 6 | na | 35.83 | 0.04 | 56.98 | 0.36 | 7.33 | 100.5 | na | 49.24 | 0.03 | 44.96 | 0.27 | 5.50 | 100 | 0.91 | 1.03 |
| CM2 | PPI | QUE97_C6S2 | 2 | na | 35.82 | 0.06 | 59.12 | 0.23 | 4.59 | 99.8 | na | 49.46 | 0.05 | 46.86 | 0.17 | 3.46 | 100 | 0.95 | 1.02 |
| CM2 | PPI | QUE97_C6S2 | 3 | na | 35.63 | 0.05 | 58.85 | 0.14 | 4.06 | 98.7 | na | 49.66 | 0.05 | 47.09 | 0.11 | 3.09 | 100 | 0.95 | 1.01 |
| CM2 | PPI | QUE97_C12S2 | 1 | na | 36.63 | 0.02 | 62.01 | 0.19 | 1.25 | 100.1 | na | 50.16 | 0.02 | 48.74 | 0.14 | 0.94 | 100 | 0.97 | 0.99 |
| CM2 | PPI | QUE97_C12S2 | 2 | na | 36.95 | 0.03 | 61.76 | 0.19 | 1.37 | 100.3 | na | 50.43 | 0.02 | 48.39 | 0.14 | 1.02 | 100 | 0.96 | 0.98 |
| CM2 | PPI | QUE97_C12S2 | 3 | na | 36.78 | 0.02 | 61.57 | 0.21 | 1.42 | 100.0 | na | 50.37 | 0.02 | 48.40 | 0.16 | 1.06 | 100 | 0.96 | 0.99 |
| CM2 | PPI | QUE97_C12S2 | 4 | na | 36.77 | 0.03 | 62.00 | 0.17 | 1.10 | 100.1 | na | 50.31 | 0.03 | 48.71 | 0.12 | 0.82 | 100 | 0.97 | 0.99 |
| CM2 | PPI | QUE97_C12S2 | 5 | na | 36.87 | 0.04 | 61.59 | 0.20 | 1.46 | 100.2 | na | 50.40 | 0.04 | 48.33 | 0.15 | 1.09 | 100 | 0.96 | 0.98 |
| CM2 | PPI | QUE97_C12S2 | 6 | na | 36.53 | 0.05 | 61.82 | 0.19 | 1.01 | 99.6 | na | 50.24 | 0.04 | 48.82 | 0.14 | 0.76 | 100 | 0.97 | 0.99 |
| CM2 | PPI | QUE97_C12S2 | 7 | na | 36.83 | 0.05 | 61.87 | 0.17 | 1.28 | 100.2 | na | 50.34 | 0.04 | 48.54 | 0.13 | 0.95 | 100 | 0.96 | 0.99 |
| CM2 | PPI | QUE97_C12S3 | 1 | na | 36.58 | 0.02 | 61.46 | 0.19 | 1.47 | 99.7 | na | 50.26 | 0.02 | 48.48 | 0.14 | 1.10 | 100 | 0.96 | 0.99 |
| CM2 | PPI | QUE97_C12S3 | 2 | na | 36.94 | 0.02 | 60.63 | 0.18 | 1.75 | 99.5 | na | 50.73 | 0.02 | 47.80 | 0.13 | 1.31 | 100 | 0.94 | 0.97 |
| CM2 | PPI | QUE97_C12S3 | 3 | na | 36.90 | 0.02 | 61.46 | 0.22 | 1.70 | 100.3 | na | 50.38 | 0.02 | 48.17 | 0.16 | 1.27 | 100 | 0.96 | 0.98 |
| CM2 | PPI | QUE97_C12S3 | 4 | na | 36.72 | 0.03 | 60.86 | 0.22 | 1.85 | 99.7 | na | 50.44 | 0.02 | 47.99 | 0.17 | 1.39 | 100 | 0.95 | 0.98 |
| CM2 | PPI | QUE97_C12S3 | 5 | na | 37.10 | 0.02 | 60.99 | 0.24 | 1.90 | 100.3 | na | 50.62 | 0.02 | 47.77 | 0.18 | 1.42 | 100 | 0.94 | 0.98 |
| CM2 | PPI | QUE97_C12S3 | 6 | na | 37.27 | 0.03 | 61.50 | 0.25 | 1.72 | 100.8 | na | 50.59 | 0.02 | 47.93 | 0.19 | 1.28 | 100 | 0.95 | 0.98 |
| CM2 | PPI | QUE97_C12S3 | 7 | na | 37.04 | 0.03 | 61.21 | 0.21 | 1.75 | 100.2 | na | 50.56 | 0.02 | 47.96 | 0.15 | 1.30 | 100 | 0.95 | 0.98 |
| CM2 | PPI | QUE97_C18S2 | 1 | na | 36.60 | 0.04 | 62.93 | 0.09 | 0.84 | 100.5 | na | 49.97 | 0.03 | 49.31 | 0.06 | 0.62 | 100 | 0.99 | 1.00 |
| CM2 | PPI | QUE97_C18S2 | 2 | na | 36.70 | 0.05 | 62.83 | 0.08 | 0.89 | 100.6 | na | 50.05 | 0.04 | 49.18 | 0.06 | 0.66 | 100 | 0.98 | 1.00 |
| CM2 | PPI | QUE97_C18S2 | 3 | na | 36.84 | 0.04 | 63.31 | 0.09 | 0.50 | 100.8 | na | 50.10 | 0.03 | 49.43 | 0.06 | 0.37 | 100 | 0.99 | 1.00 |
| CM2 | PPI | QUE97_C18S2 | 4 | na | 36.92 | 0.05 | 63.17 | 0.07 | 0.44 | 100.6 | na | 50.24 | 0.04 | 49.34 | 0.05 | 0.33 | 100 | 0.98 | 0.99 |
| CM2 | PPI | QUE97_C18S2 | 5 | na | 36.71 | 0.03 | 63.05 | 0.10 | 0.53 | 100.4 | na | 50.11 | 0.03 | 49.40 | 0.08 | 0.39 | 100 | 0.99 | 1.00 |
| CM2 | PPI | QUE97_C18S2 | 6 | na | 37.01 | 0.11 | 62.54 | 0.10 | 0.56 | 100.3 | na | 50.46 | 0.10 | 48.95 | 0.07 | 0.42 | 100 | 0.97 | 0.98 |
| CM2 | PPI | QUE97_C18S2 | 7 | na | 36.74 | 0.06 | 62.65 | 0.08 | 0.43 | 100.0 | na | 50.32 | 0.05 | 49.25 | 0.06 | 0.32 | 100 | 0.98 | 0.99 |
| CM2 | PPI | QUE97_C18S2 | 8 | na | 36.12 | 0.05 | 62.96 | 0.06 | 0.50 | 99.7 | na | 49.75 | 0.04 | 49.79 | 0.04 | 0.38 | 100 | 1.00 | 1.01 |
| CM2 | PPI | QUE97_C18S2 | 12 | na | 36.78 | 0.06 | 63.06 | 0.09 | 0.41 | 100.4 | na | 50.19 | 0.05 | 49.40 | 0.06 | 0.30 | 100 | 0.98 | 0.99 |
| CM2 | PPI | QUE97_S1 | 1 | na | 37.06 | 0.03 | 61.41 | 0.15 | 1.51 | 100.2 | na | 50.60 | 0.02 | 48.14 | 0.11 | 1.12 | 100 | 0.95 | 0.98 |
| CM2 | PPI | QUE97_S1 | 2 | na | 37.58 | 0.02 | 60.90 | 0.14 | 1.65 | 100.3 | na | 51.11 | 0.02 | 47.54 | 0.10 | 1.22 | 100 | 0.93 | 0.96 |
| CM2 | PPI | QUE97_S1 | 3 | na | 37.59 | 0.02 | 60.82 | 0.20 | 1.66 | 100.3 | na | 51.12 | 0.02 | 47.48 | 0.15 | 1.23 | 100 | 0.93 | 0.96 |
| CM2 | PPI | QUE97_S1 | 4 | na | 37.47 | 0.02 | 61.34 | 0.15 | 1.74 | 100.7 | na | 50.82 | 0.01 | 47.76 | 0.11 | 1.29 | 100 | 0.94 | 0.97 |
| CM2 | PPI | QUE97_S1 | 5 | na | 37.58 | 0.02 | 60.75 | 0.21 | 1.85 | 100.4 | na | 51.06 | 0.02 | 47.39 | 0.16 | 1.37 | 100 | 0.93 | 0.96 |
| CM2 | PPI | QUE97_S1 | 6 | na | 37.42 | 0.01 | 60.75 | 0.22 | 1.82 | 100.2 | na | 50.97 | 0.01 | 47.50 | 0.16 | 1.36 | 100 | 0.93 | 0.96 |
| CM2 | PPI | QUE97_S1 | 7 | na | 37.45 | 0.03 | 60.48 | 0.22 | 2.03 | 100.2 | na | 51.01 | 0.02 | 47.29 | 0.17 | 1.51 | 100 | 0.93 | 0.96 |
| CM2 | PPI | QUE97_S1 | 8 | na | 37.44 | 0.02 | 60.53 | 0.22 | 2.25 | 100.5 | na | 50.90 | 0.02 | 47.24 | 0.17 | 1.67 | 100 | 0.93 | 0.96 |
| CM2 | PPI | QUE97_S1 | 9 | na | 37.57 | 0.06 | 60.99 | 0.21 | 1.75 | 100.6 | na | 50.99 | 0.05 | 47.51 | 0.16 | 1.30 | 100 | 0.93 | 0.96 |

Table A5 cont.

| Meteorite group | Textural group | Grain | Point | Weight % | | | | | | | Atomic % | | | | | | | Atomic ratio | |
|---|---|---|---|---|---|---|---|---|---|---|---|---|---|---|---|---|---|---|---|
| | | | | P | S | Cr | Fe | Co | Ni | Total | P | S | Cr | Fe | Co | Ni | Total | Fe/S | (Fe+Co+Ni+Cr)/S |
| CM2 | PPI | QUE97_S1 | 10 | na | 37.66 | 0.02 | 60.65 | 0.21 | 2.02 | 100.6 | na | 51.10 | 0.02 | 47.23 | 0.16 | 1.49 | 100 | 0.92 | 0.96 |
| CM2 | PPI | QUE97_S4 | 1 | na | 36.27 | 0.03 | 61.96 | 0.20 | 1.23 | 99.7 | na | 49.93 | 0.02 | 48.97 | 0.15 | 0.93 | 100 | 0.98 | 1.00 |
| CM2 | PPI | QUE97_S4 | 2 | na | 37.08 | 0.01 | 62.21 | 0.21 | 1.31 | 100.8 | na | 50.36 | 0.00 | 48.50 | 0.15 | 0.97 | 100 | 0.96 | 0.99 |
| CM2 | PPI | QUE97_S4 | 3 | na | 37.03 | 0.02 | 62.19 | 0.20 | 1.23 | 100.7 | na | 50.37 | 0.01 | 48.56 | 0.15 | 0.92 | 100 | 0.96 | 0.99 |
| CM2 | PPI | QUE97_S4 | 4 | na | 37.27 | 0.02 | 62.37 | 0.18 | 1.11 | 101.0 | na | 50.51 | 0.02 | 48.52 | 0.13 | 0.82 | 100 | 0.96 | 0.98 |
| CM2 | PPI | QUE97_S4 | 5 | na | 37.07 | 0.02 | 62.47 | 0.19 | 1.09 | 100.8 | na | 50.34 | 0.01 | 48.70 | 0.14 | 0.81 | 100 | 0.97 | 0.99 |
| CM2 | PPI | QUE97_S4 | 6 | na | 37.00 | 0.01 | 61.34 | 0.17 | 2.04 | 100.6 | na | 50.39 | 0.00 | 47.96 | 0.13 | 1.52 | 100 | 0.95 | 0.98 |
| CM2 | PPI | QUE97_S4 | 7 | na | 37.27 | 0.00 | 62.08 | 0.14 | 1.07 | 100.6 | na | 50.66 | 0.00 | 48.44 | 0.10 | 0.80 | 100 | 0.96 | 0.97 |
| CM2 | PPI | QUE97_S4 | 8 | na | 36.99 | 0.01 | 62.48 | 0.18 | 0.91 | 100.6 | na | 50.36 | 0.01 | 48.82 | 0.13 | 0.68 | 100 | 0.97 | 0.99 |
| CM2 | PPI | QUE97_S4 | 9 | na | 36.62 | 0.01 | 62.30 | 0.17 | 1.03 | 100.1 | na | 50.13 | 0.01 | 48.96 | 0.12 | 0.77 | 100 | 0.98 | 0.99 |
| CM2 | PPI | QUE97_S4 | 10 | na | 36.89 | 0.00 | 61.09 | 0.20 | 2.26 | 100.5 | na | 50.33 | 0.00 | 47.84 | 0.15 | 1.68 | 100 | 0.95 | 0.99 |
| CM2 | PPI | QUE97_S6 | 1 | na | 37.05 | 0.13 | 58.98 | 0.23 | 3.13 | 99.5 | na | 50.88 | 0.11 | 46.49 | 0.17 | 2.35 | 100 | 0.91 | 0.97 |
| CM2 | PPI | QUE97_S6 | 2 | na | 36.96 | 0.14 | 59.58 | 0.15 | 2.90 | 99.7 | na | 50.69 | 0.12 | 46.91 | 0.11 | 2.17 | 100 | 0.93 | 0.97 |
| CM2 | PPI | QUE97_S6 | 3 | na | 36.89 | 0.14 | 58.80 | 0.23 | 3.28 | 99.3 | na | 50.79 | 0.12 | 46.46 | 0.17 | 2.47 | 100 | 0.91 | 0.97 |
| CM2 | PPI | QUE97_S6 | 4 | na | 36.81 | 0.14 | 59.53 | 0.18 | 2.83 | 99.5 | na | 50.63 | 0.12 | 46.99 | 0.14 | 2.13 | 100 | 0.93 | 0.98 |
| CM2 | PPI | QUE97_S6 | 5 | na | 36.94 | 0.13 | 59.38 | 0.16 | 2.98 | 99.6 | na | 50.72 | 0.11 | 46.81 | 0.12 | 2.24 | 100 | 0.92 | 0.97 |
| CM2 | PPI | QUE97_S6 | 6 | na | 37.14 | 0.14 | 58.70 | 0.24 | 3.62 | 99.8 | na | 50.86 | 0.12 | 46.14 | 0.18 | 2.70 | 100 | 0.91 | 0.97 |
| CM2 | PPI | QUE97_S6 | 7 | na | 37.20 | 0.14 | 59.77 | 0.18 | 2.95 | 100.2 | na | 50.75 | 0.12 | 46.81 | 0.13 | 2.20 | 100 | 0.92 | 0.97 |
| CM2 | PPI | QUE97_S6 | 8 | na | 37.29 | 0.14 | 59.59 | 0.13 | 2.92 | 100.1 | na | 50.91 | 0.12 | 46.70 | 0.10 | 2.18 | 100 | 0.92 | 0.96 |
| CM2 | PPI | QUE97_S6 | 9 | na | 36.80 | 0.14 | 59.75 | 0.08 | 2.60 | 99.4 | na | 50.66 | 0.11 | 47.21 | 0.06 | 1.96 | 100 | 0.93 | 0.97 |
| CM2 | PPI | QUE97_S12 | 1 | na | 36.07 | 0.03 | 61.71 | 0.24 | 1.43 | 99.5 | na | 49.81 | 0.02 | 48.91 | 0.18 | 1.08 | 100 | 0.98 | 1.01 |
| CM2 | PPI | QUE97_S12 | 2 | na | 35.95 | 0.03 | 61.78 | 0.20 | 1.44 | 99.4 | na | 49.71 | 0.02 | 49.03 | 0.15 | 1.09 | 100 | 0.99 | 1.01 |
| CM2 | PPI | QUE97_S12 | 3 | na | 36.16 | 0.03 | 61.76 | 0.20 | 1.38 | 99.5 | na | 49.88 | 0.02 | 48.91 | 0.15 | 1.04 | 100 | 0.98 | 1.00 |
| CM2 | PPI | QUE97_S12 | 4 | na | 36.02 | 0.04 | 61.79 | 0.21 | 1.34 | 99.4 | na | 49.78 | 0.03 | 49.02 | 0.16 | 1.01 | 100 | 0.98 | 1.01 |
| CM2 | PPI | QUE97_S12 | 5 | na | 35.60 | 0.03 | 61.70 | 0.17 | 1.43 | 98.9 | na | 49.50 | 0.03 | 49.25 | 0.13 | 1.08 | 100 | 0.99 | 1.02 |
| CM2 | PPI | Murchison_C5S2 | 1 | na | 35.80 | 0.02 | 61.50 | 0.00 | 0.12 | 97.4 | na | 50.29 | 0.01 | 49.60 | 0.00 | 0.09 | 100 | 0.99 | 0.99 |
| CM2 | PPI | Murchison_C7S1 | 1 | na | 36.10 | 0.03 | 61.75 | 0.13 | 0.50 | 98.7 | na | 50.21 | 0.02 | 49.30 | 0.10 | 0.38 | 100 | 0.98 | 0.99 |
| CM2 | PPI | Murchison_C7S1 | 2 | na | 36.38 | 0.02 | 61.48 | 0.19 | 0.86 | 99.0 | na | 50.35 | 0.02 | 48.84 | 0.14 | 0.65 | 100 | 0.97 | 0.99 |
| CM2 | PPI | Murchison_C7S1 | 5 | na | 36.32 | 0.02 | 60.32 | 0.15 | 1.80 | 98.6 | na | 50.43 | 0.02 | 48.07 | 0.11 | 1.37 | 100 | 0.95 | 0.98 |
| CM2 | PPI | Murchison_S2 | 1 | na | 36.77 | 0.01 | 60.24 | 0.21 | 0.96 | 98.3 | na | 51.08 | 0.01 | 48.03 | 0.16 | 0.73 | 100 | 0.94 | 0.96 |
| CM2 | PPI | Murchison_S2 | 2 | na | 36.35 | 0.01 | 60.18 | 0.19 | 0.91 | 97.7 | na | 50.84 | 0.01 | 48.31 | 0.14 | 0.69 | 100 | 0.95 | 0.97 |
| CM2 | PPI | Murchison_S4 | 1 | na | 36.52 | 0.02 | 61.68 | 0.11 | 0.52 | 98.9 | na | 50.52 | 0.02 | 48.99 | 0.08 | 0.39 | 100 | 0.97 | 0.98 |
| CM2 | PPI | Murchison_S4 | 2 | na | 36.58 | 0.02 | 61.33 | 0.09 | 0.55 | 98.6 | na | 50.70 | 0.02 | 48.80 | 0.07 | 0.41 | 100 | 0.96 | 0.97 |
| CM2 | PPI | Murchison_S4 | 3 | na | 36.54 | 0.03 | 61.22 | 0.10 | 0.56 | 98.5 | na | 50.70 | 0.03 | 48.77 | 0.08 | 0.43 | 100 | 0.96 | 0.97 |
| CM2 | PPI | Murchison_S4 | 4 | na | 36.11 | 0.04 | 61.61 | 0.09 | 0.44 | 98.3 | na | 50.30 | 0.03 | 49.27 | 0.07 | 0.33 | 100 | 0.98 | 0.99 |

**Table A5 cont.**

| Meteorite group | Textural group | Grain | Point | Weight % | | | | | | | Atomic % | | | | | | | Atomic ratio | |
|---|---|---|---|---|---|---|---|---|---|---|---|---|---|---|---|---|---|---|---|
| | | | | P | S | Cr | Fe | Co | Ni | Total | P | S | Cr | Fe | Co | Ni | Total | Fe/S | (Fe+Co+Ni+Cr)/S |
| CM2 | PPI | Murchison_S4 | 5 | na | 36.28 | 0.01 | 61.42 | 0.05 | 0.44 | 98.2 | na | 50.52 | 0.01 | 49.10 | 0.04 | 0.33 | 100 | 0.97 | 0.98 |
| CM2 | PPI | Murray_C3S1 | 2 | na | 36.31 | na | 60.59 | 0.13 | 0.64 | 97.8 | na | 50.77 | 0.00 | 48.64 | 0.10 | 0.49 | 100 | 0.96 | 0.97 |
| CM2 | PPI | Murray_S6 | 1 | na | 35.55 | na | 60.47 | 0.23 | 1.25 | 97.6 | na | 50.02 | 0.00 | 48.84 | 0.18 | 0.96 | 100 | 0.98 | 1.00 |
| CM2 | PPI | Murray_S6 | 2 | na | 35.77 | na | 61.62 | 0.16 | 0.84 | 98.4 | na | 49.90 | 0.00 | 49.34 | 0.12 | 0.64 | 100 | 0.99 | 1.00 |
| CM2 | PPI | Murray_S6 | 3 | na | 35.67 | na | 61.42 | 0.22 | 0.89 | 98.3 | na | 49.87 | 0.00 | 49.28 | 0.17 | 0.68 | 100 | 0.99 | 1.01 |
| CM2 | PPI | Murray_S6 | 7 | na | 35.27 | na | 61.09 | 0.15 | 0.79 | 97.3 | na | 49.78 | 0.00 | 49.50 | 0.11 | 0.61 | 100 | 0.99 | 1.01 |
| CM2 | PPI | Murray_S10 | 1 | na | 35.78 | na | 60.33 | 0.20 | 1.46 | 97.8 | na | 50.17 | 0.00 | 48.56 | 0.15 | 1.12 | 100 | 0.97 | 0.99 |
| CM2 | PPI | Murray_S10 | 2 | na | 35.82 | na | 60.21 | 0.19 | 1.37 | 97.7 | na | 50.28 | 0.00 | 48.52 | 0.15 | 1.05 | 100 | 0.96 | 0.99 |
| CM2 | PPI | Murray_S13 | 1 | na | 35.66 | na | 59.22 | 0.37 | 2.42 | 97.7 | na | 50.10 | 0.00 | 47.76 | 0.28 | 1.86 | 100 | 0.95 | 1.00 |
| CM2 | PPI | Murray_S13 | 2 | na | 35.74 | na | 59.67 | 0.31 | 2.16 | 97.9 | na | 50.09 | 0.00 | 48.02 | 0.24 | 1.65 | 100 | 0.96 | 1.00 |
| CM2 | PPI | Murray_S13 | 3 | na | 35.51 | na | 60.34 | 0.25 | 1.26 | 97.4 | na | 50.03 | 0.00 | 48.81 | 0.19 | 0.97 | 100 | 0.98 | 1.00 |
| CM2 | PPI | Murray_S13 | 4 | na | 35.42 | na | 60.79 | 0.15 | 0.90 | 97.3 | na | 49.97 | 0.00 | 49.23 | 0.11 | 0.69 | 100 | 0.99 | 1.00 |
| CM2 | PPI | Murray_S13 | 5 | na | 35.79 | na | 61.39 | 0.09 | 0.38 | 97.6 | na | 50.21 | 0.00 | 49.44 | 0.07 | 0.29 | 100 | 0.98 | 0.99 |

[1]CM2 data is from Singerling & Brearley (2018) and (2020).

PPM = pyrrhotite-pentlandite-magnetite, PPI alt mgt = pyrrhotite-pentlandite intergrowth grains altering to magnetite, PPI = pyrrhotite-pentlandite intergrowth grains, ALH49 = ALH 84049, LAP = LAP 031166, QUE97 = QUE 97990, S# = matrix sulfide #, C#S# = sulfide # in chondrule #, bdl = below detection limit, na = not analyzed.